\documentclass{aa}  
\usepackage{graphicx}
\usepackage{txfonts}
\begin{document}
   \title{Dynamical properties of a large young disk galaxy at z=2.03\thanks{Observations carried out using the Very Large Telescope at the ESO Paranal Observatory, Program ID 077.B-0079.} }

   \subtitle{   }

   \author{L. van Starkenburg \inst{1}
          \and
          P. P. van der Werf \inst{1}
          \and
          M. Franx \inst{1}
          \and
          I. Labb\'e\inst{2}
          \and
          G. Rudnick\inst{3}
          \and 
          S. Wuyts\inst{4}
           }

   \offprints{}

   \institute{Leiden Observatory, Leiden University, P.O. Box 9513, 2300 RA  Leiden, The Netherlands \\
              \email{vstarken, pvdwerf, franx@strw.leidenuniv.nl }
         \and Carnegie Observatories, Carnegie Institution of Washington, 813 Santa Barbara Street, Pasadena, CA 91101, USA \\
             \email{ivo@ociw.edu}
         \and Goldberg Fellow, National Optical Astronomy Observatory, Tucson, AZ 85719, USA\\
            \email{grudnick@noao.edu}
          \and Harvard-Smithsonian Center for Astrophysics, 60 Garden St., Cambridge, MA 02138, USA\\
             \email{swuyts@cfa.harvard.edu}
            }

   \date{Received ; accepted }

  \abstract
   {The study of high redshift Tully-Fisher relations (TFRs) is limited by the use of long slit spectrographs, rest frame B band and star formation selected galaxies.  } 
   {We try to circumvent these issues by using integral field spectroscopy (SINFONI), by studying the rest frame K band and stellar mass TFR, and by selecting targets without a bias to strongly star forming galaxies. In this paper, we demonstrate our methods on our best case. This galaxy, F257, at $z=2.03$, was selecte from a sample of  candidate high redshift large disk galaxies in the Hubble Deep Field South that were selected with photometric and morphological criteria. }
   {We used SINFONI at the VLT to obtain an integral field spectrum of the H$\alpha$ line and hence a velocity field and rotation curve. We also use UBVIJHK+IRAC band photometry to determine a stellar photometric mass.    }
   {We find that F257 is indistinguishable from local late type galaxies in many respects: it has a regular velocity field, increasing velocity disperion towards its center, its rotation curve flattens at 1-2 disk scale lengths, it has the same specific angular momentum as local disks, its properties are consistent with the local K band TFR. Although mainly rotationally supported, its gas component is  dynamically heated with respect to local galaxies ($V/\sigma_{z}\sim4$) and it is offset from the local stellar mass TFR at the $2\sigma$ level.  But, this offset depends on the SED modeling parameters. In particular, for a 2-component star formation history (SFH), F257 is in agreement with the local stellar mass TFR. F257 is then a nearly ($\sim 75\%$) maximum disk. The dynamical properties of F257 are more like those of local galaxies than those of any other galaxy at similar redshift observed to date. However, the gas-to-stellar mass ratio is unusally large: 2.5. }
    {}
 
   \maketitle
%

\section{Introduction}

Starting with density fluctuations in the early universe, dark matter halos grow through gravity. Galaxy formation is described by semi-analytical models or hydrodynamical simulations. These may include physical processes such as gas cooling, star formation, feedback from stellar winds, supernovae and/or AGN, chemical evolution and stellar populations (Okamoto \cite{okamoto} and references therein). 

 The disk formation epoch in earlier generations of models is relatively late and mostly at $z \lesssim 1$. These early models also predicted that these disks are much smaller than observed in local spiral galaxies and  that their specific angular momenta are only  20\% of that of their halo's (Navarro, Frenk \& White \cite{navarro}). More recent models produce more realistic size disks (Mo et al. \cite{mo1998}, Portinari \& Sommer-Larsen \cite{portinari}, Governato et al. \cite{governato}) and place the formation epoch somewhat earlier: $0.4<z<2.0$ (Buchalter, Jimenez \& Kamionkowski \cite{buchalter}). Reproducing the specific angular momentum of local disk galaxies remains a challenge (e.g. Portinari \& Sommer-Larsen \cite{portinari}).

More than 30 years after the first publication of what is now known as the Tully-Fisher relation (TFR) (Tully \& Fisher \cite{tullyfisher}), the origin of the TFR is still subject of debate. Some authors explain the TFR as a consequence of self-regulated star formation in disks with different masses (e.g. Silk \cite{silk}). In other models, the TFR is a direct consequence of the cosmological equivalence between mass and circular velocity resulting from the finite age of the universe which imposes a maximum radius from with matter can accrete to form a disk (e.g. Mo et al. \cite{mo1998}, Steinmetz \& Navarro \cite{steinmetz}). Steinmetz \& Navarro (\cite{steinmetz}) show that although the slope of the TFR can be naturally explained by hierarchical models, the normalization and evolution of the TFR depend strongly on the star formation prescriptions and the cosmological parameters. Recently, Courteau et al. (\cite{courteau}) showed that two long known, different theoretical predictions for two different slopes of the TFR are in fact related.

Disk galaxy scaling relations, like the TFR, put strong constraints on galaxy formation scenarios, and in particular on the connection of the dark and baryonic component (van den Bosch \cite{bosch}). The zero-point of the TFR constrains the SFH and the cosmological parameters that control the universal baryon fraction and the time of assembly of galaxies of different mass (Steinmetz \& Navarro \cite{steinmetz}). Many authors struggle to reproduce the zero-point of the TFR (e.g. Portinari \& Sommer-Larsen \cite{portinari}, Steinmetz \& Navarro \cite{steinmetz}). The slope of the TFR and the steepening of the slope towards longer wavelengths can only be explained by supernova feedback (van den Bosch \cite{bosch}) although others claim it is a natural result of the hierarchical models (Steinmetz \& Navarro \cite{steinmetz}). The small scatter in the TFR is explained by scatter in halo spin parameters and formation redshifts (e.g. van den Bosch \cite{bosch}).   Even the most recent galaxy formation and evolution models are not able to simultaneously reproduce for example the slope, zero-point and scatter of the TFR and the zero-point and shape of the luminosity function (e.g. Courteau et al. \cite{courteau} and references therein).

The TFR defines a tight correlation between the luminosity and the flat rotation curve velocity of spiral galaxies. 
If luminosity is converted into stellar mass (which requires detailed modelling of the stellar spectral energy distribution (SED)), the tight correlation remains: the stellar mass TFR. 
Semi-analytic models predict evolution in the slope, zero-point and scatter properties of the TFR and the stellar mass TFR (e.g. Steinmetz \& Navarro \cite{steinmetz}, Buchalter, Jimenez \& Kamionkowski \cite{buchalter}, Portinari \& Sommer-Larsen \cite{portinari}). 
A key element is the relative evolution of the  dark and stellar mass components, which is a direct test of the stellar mass buildup within a dark halo of a given mass.

Critical tests of these theories can be made by observing the evolution of disks backward in time. Rotation curves (RCs) of field galaxies out to $z = 1$ measured with the Keck telescope and Very Large Telescope (VLT) are found to be similar to those of nearby galaxies in both shape and amplitude implying that some massive disks were already in place at $z \sim 1$ (Vogt et al. \cite{vogt1996}, \cite{vogt1997}, Barden et al. \cite{barden}, Swinbank et al. \cite{swinbank}). Observations of $z \sim 2$ galaxies have revealed ordered rotation, but only the rising part of the rotation curve (e.g. the lensed sources of Lemoine-Busserolle et al. \cite{lemoine}). Van Starkenburg et al. (\cite{vanstarkenburg}) found ordered rotation in a $z = 2.46$ galaxy, with an asymmetric line profile indicating a central starburst and weaker starformation in the outer parts. Ordered rotation has been found out to $z \sim 3.2$ by Moorwood et al. (\cite{moorwood00}, \cite{moorwood03}) and by Nesvadba et al. (\cite{nesvadba6}), the latter again on a lensed source.

Also, several attempts to measure size evolution and the TFR at redshift $0.5 - 1.5$ have been made over the past few years with contradictory claims about disk evolution. For instance, Mo et al. (\cite{mo1998}) used previously acquired data (Vogt et al.\cite{vogt1996}, \cite{vogt1997}; Schade et al. \cite{schade}; Lilly et al. \cite{lilly}) to conclude that $z \sim 1 $  galaxies are smaller than local galaxies at a fixed rest-frame B-magnitude or rotation velocity, consistent with hierarchical models. A brightening with redshift of disk galaxies at fixed rotation speeds was also detected by Barden et al. (\cite{barden}). However, Simard et al. (\cite{simard}) and Ziegler et al. (\cite{ziegler}) have shown that massive galaxies at $z < 1$ are consistent with having no evolution in surface brightness, once selection effects are taken into account. 

The largest source of uncertainty in those studies, all done with long slit spectra, are the velocity measurements. In recent years, it has become possible to do integral field spectroscopy at near infrared wavelengths, for example with SINFONI at the VLT (Eisenhauer et al. \cite{eisenhauer}, Bonnet et al. \cite{bonnet}). Integral field spectroscopy has significant advantages over long slit spectroscopy. First, long slit measurements may suffer from slit misalignment. More fundamentally, the photometric and kinematic axes may be different (due to star forming regions or intrinsic shape of galaxy). Kinematic disturbances are harder or impossible to identify in slit spectra. Integral field spectroscopy does not have all these slit related problems. Finally, integral field spectroscopy makes for a fairer comparison with low-$z$ Tully-Fisher data, which are typically derived from full imaging of the HI velocity field with radio aperture synthesis. Integral field spectroscopy thus presents a major improvement over slit spectroscopy, provided significant resolution over the rotating disk is achieved. Recently, F\"orster Schreiber et al. (\cite{forster}) and Bouch\'e et al. (\cite{bouche}) used SINFONI to obtain two-dimensional velocity fields of a number of galaxies at $z \sim 2$ with different selection criteria, clearly demonstrating the power of an efficient integral-field spectrograph such as SINFONI for this type of work.

A second major problem that plagued previous high-$z$ TFR studies is the use of rest-frame B-band, which is very sensitive to small amounts of recent star formation and extinction by dust. There are now a number of deep surveys available with comprehensive photometric coverage ideal for studying the high redshift TFR in rest frame near-IR luminosities. We used the HDF-S and MS1054 fields which includes recent Spitzer IRAC data (Labb\'e et al. \cite{irac}). This allows us to use rest-frame near-IR luminosities, which define much tighter TFRs than rest-frame optical luminosities (e.g. Verheijen \cite{verheijen}). Rest-frame K band luminosities as derived from Spitzer data are much less sensitive to the effects of dust and recent starformation than rest frame optical or UV luminosities and will provide us with more accurate stellar masses.

A third limitation of previous studies is that many of these target starformation-selected galaxies (e.g., selected on rest-frame UV or bright emission lines). This compromises attempts to use these galaxies to trace the evolution of the TFR (see van Starkenburg et al. \cite{vanstarkenburg} for a detailed discussion), since local TFR samples are composed of relatively quiescent galaxies. Although a certain amount of star formation is necessary to observe the velocity fields of high redshift galaxies, there are several ways to partly circumvent this problem.

First, by selecting from ultra-deep near-IR imaging surveys, such as the FIRES data of the HDF-S and MS1054 fields\footnote{The FIRES data are publicly available, see http://www.strw.leidenuniv.nl/\~{}fires}(Labb\'e et al. \cite{fireshdfs}, Franx et al. \cite{franx}, F\"orster-Schreiber et al. \cite{firesms1054}), we can select galaxies in the rest frame optical instead of the rest frame UV. The FIRES data is sufficiently deep to select galaxies in the rest-frame optical out to $z \sim 4$ and the depth on the Hubble Deep Field South (HDF-S) is sufficient to detect non-evolving ($0.6L_*$  at $z = 0$) spirals out to $z = 3$.

A second improvement is to preselect massive galaxies. Contrary to low mass galaxies, their $\log(L)$ (and hence offset from the TFR) is not or less affected by the luminosity of  a starburst  (van Starkenburg et al. \cite{vanstarkenburg}). We observed several targets in this category in two redshift ranges, which will be discussed in two separate papers (van Starkenburg et al. \cite{prep1}, \cite{prep2}).

Finally, it is now possible to select high redshift galaxies that morphologically resemble local disks. A search for such disks in the FIRE survey of the  2\farcm5 x 2\farcm5 HDF-S field resulted in a surprisingly large number (6) of candidate large disks (Labb\'e et al. \cite{labbe}). These galaxies, at redshifts 1.4 to 3.0 (most of them spectroscopically confirmed), have exponential profiles with effective radii ranging from 0\farcs65 to 0\farcs9 (or $5.0 - 7.5 h^{-1}_{70}$ kpc) and extending over 2-3 effective radii and appear remarkably regular in the near infrared (rest frame optical). Four are well fit by just exponential disks and two require an additional $r^{1/4}$  bulge  component, contributing about 40\% of the light. At shorter wavelengths, they have more extended and irregular morphologies. They have strong rest-frame optical breaks in their centres and evidence for intense star formation in the outer parts, reminiscent of local spirals, with blue star-forming disks and red bulges. Their sizes are comparable to the Milky Way and those of L* disks in the local universe. Specifically, these sizes are much larger than those of typical LBGs ($1 - 2 h^{-1}_{70}$ kpc; Giavalisco et al. \cite{giavalisco}; Lowenthal et al. \cite{lowenthal})) and their sizes are in the upper end of the range found by Trujillo et al. (\cite{trujillo}) for all galaxies at similar $z$ in the two FIRES fields. They are bright in K (K$ = 19.7 -  21.1$) and have large stellar masses ($M > 10^{11}\textrm{ M}_\odot$). They constitute half of the most rest-frame luminous galaxies ($L_V > 6 \times 10^{10} h^{-2}_{70} L_\odot $) and their number density exceeds model predictions (e.g., Mo et al., \cite{mo1998}) by a factor of two.  
Obviously, the first and foremost question that needs to be resolved is whether these objects are truly rotationally supported disks, and this can only be tested by obtaining kinematic data.

In this paper, we present SINFONI observations of one of the large disks of the Labb\'e et al. sample (id 257 in the FIRES HDF-S catalog). These observations were part of a larger project of SINFONI observations of mass and/or disk selected high redshift galaxies, which will be presented in two separate papers (van Starkenburg et al. \cite{prep1}, \cite{prep2}). Here, we will discuss our methods as used on our best case.

\begin{figure}

\includegraphics[width=8cm]{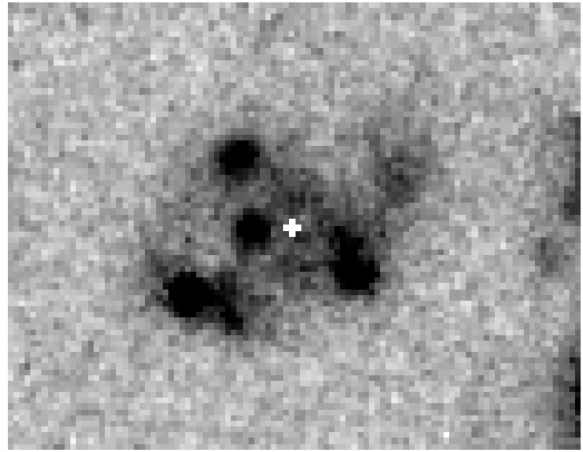}
\includegraphics[width=8cm]{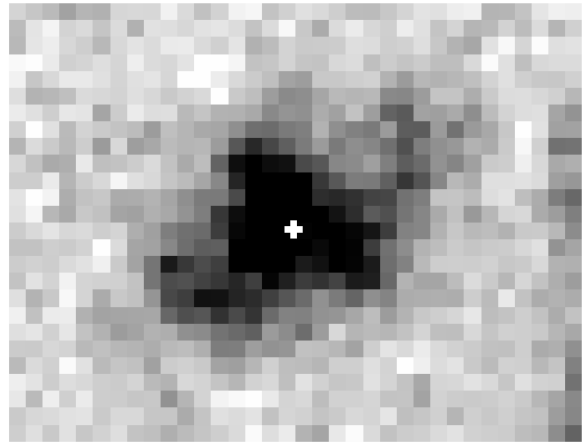}
\includegraphics[width=8cm]{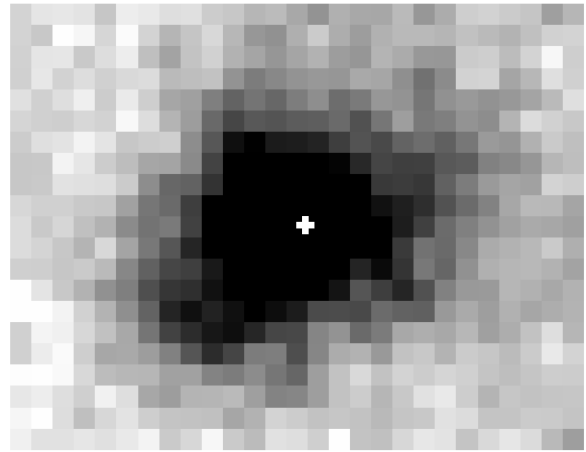}

 \caption{From top to bottom: F814W image, NICMOS H band image and Ks image from FIRES data. Image size is always 3\arcsec times 4\arcsec and the images have been rotated so that east is to the top and north is to the right. The (K band) center from the FIRES data is marked by the white crosses.} 
 \label{images}

\end{figure}

We observed F257 with SINFONI during the night of August 15, 2006. We selected F257 as our primary target for this observing run because it is the second brightest K band source ($K=20.25$) of the large disks in the HDF-S (Labb\'e et al. \cite{labbe})  that  has a known spectroscopic redshift ($z=2.03$, Vanzella et al. \cite{vanzella}) with H$\alpha$ at an observable wavelength with SINFONI. (We had observed the brightest K band target fullfilling this criteria during an earlier observing run, see van Starkenburg et al. \cite{prep2}). Its effective radius is  0\farcs74 in K band, the total radial extent  above the K-band detection threshold is two effective radii or 3-3.5 scale lengths ($R_d=0\farcs44$). The inclination ($\sim 54\degr$, calculated from the ellipticity of Labb\'e et al. \cite{labbe}) is suited for a velocity field analysis. F257 shows a clumpy distribution in the HST F814W filter (see Fig. \ref{images}a) which corresponds to rest frame UV. The rest frame optical (observed K band) image shows a smooth exponential profile (see Fig. \ref{images}c). This difference cannot be attributed to the different PSF of the F814W and the K band image: the F814W image smoothed to the K band PSF still shows the clumps. Labb\'e et al. (\cite{labbe}) fitted an exponential profile without the need for a bulge component.  NICMOS H band shows an intermediate picture (see Fig. \ref{images}b). The S\'ersic index as determined by Trujillo et al. (\cite{trujillo}) on the FIRES (ISAAC) H band image is 0.76. Wuyts et al. (\cite{wuyts}) made an extensive study of the SEDs of high redshift galaxies in HDF-S and found for F257 $M_* = 3.5^{ +0.5}_{ -0.3} \times 10^{10} \textrm{ M}_\odot $, a mass weighted age of $\sim160 \textrm{ Myr}$ and $A_V=1.0$ using a Salpeter initial mass function (IMF), solar metallicity, a $\tau$ 300 model for the star formation history (SFH) and Bruzual \& Charlot (\cite{bruzual}) models. 
A number of $z\sim2$ galaxies have been observed with SINFONI and other IFUs (Genzel et al. \cite{genzel}, F\"orster-Schreiber et al. \cite{forster}, Bouch\'e et al. \cite{bouche}, Law et al. \cite{law}). F257 is not different in terms of colors, SFR or age, and satisfies the conventional U-dropout criterium. It is the only galaxy selected as large disk galaxy, but the other galaxies were selected from catalogs with insufficient resolution to identify morphological large disk galaxies at optical rest frame wavelengths.

Throughout this paper, we use $H_0 = 70 \textrm{ km s}^{-1}\textrm{Mpc}^{-1}$, $\Omega_M=0.3$, $\Omega_\Lambda=0.7$ and Vega magnitudes.


\section{Observing strategy}

SINFONI (Eisenhauer et al. \cite{eisenhauer}, Bonnet et al. \cite{bonnet}) is the near infrared integral field spectrograph of the VLT. SINFONI slices a square image into 32 slitlets that are combined into a single pseudo slit and then dispersed. It operates in J, H, K and H+K band and offers three image sizes. We used the 8 x 8\arcsec~ field of view (FOV) in K band. The spectral resolution in this mode is $R \sim 4500 $ or $\textrm{FWHM } \sim 75 \textrm{ km s}^{-1}$.

We used an ABA'B'-like observing cycle, comparable to the ABBA-cycles used in near-IR long slit spectroscopy. Placing our targets at different positions in the FOV instead of taking sky frames ensures optimal use of observing time. Because there is a pair of galaxies close to F257, the last two offset positions are only slightly offset from the first two positions. We rotated SINFONI's FOV so that the other galaxies do not interfere with the observations of F257 (see Appendix \ref{app} for details on SINFONI's FOV). Individual exposures were 900 s, total on source integration time 6 hours. We used 900 s exposures to keep the extra overheads (observations of PSF stars, see details in Appendix \ref{app}) within reasonable limits. As our target is too faint to acquire directly, we performed a blind offset from a bright nearby star to our target. 
The seeing varied during the night between 0\farcs35 and 0\farcs86 (measured on the PSF stars). The seeing on the weighted combined cube is 0\farcs55.
We also observed four hot standard stars for flux calibration during the entire night (which also included observation of another target).

\section{Data reduction}

We used a combination of IRAF tools and the SINFONI pipeline (version 1.3.0) (Modigliani et al. \cite{andrea}) for data reduction. We reduced the standard calibration files provided by ESO for SINFONI (dark, distortion, flat, linearity, and arc frames) using the SINFONI pipeline to get bad pixel maps, darks, flats, wavelength calibration map, the positions of the slitlets on the detector and distortion correction.

We removed bad pixels from the science images and used LA\_COSMIC (van Dokkum \cite{lacosmic}) to remove cosmic rays. Some cosmic rays close to bright OH lines remained, we removed those close to the H$\alpha$ line manually  by replacing those pixels with the interpolated value of the surrounding pixels. We noticed an odd-even effect in some slitlets, with strength varying with time and in different quarters of the detector, sometimes absent. Therefore, we averaged all columns with the mean of their right and left neighbour to remove the odd-even effect. This causes a small (5\% increase in the FWHM of the final PSF) smear in the x-direction after image reconstruction (see Appendix \ref{app} for definition of x and y direction on SINFONI's FOV).

We used the SINFONI pipeline to correct our data for distortion, flat fielding, reconstructing the data cube, and combining our science data in pairs (``first subtraction'') and did exactly the same for the PSF stars. We did an illumination correction on the science data using an early version of the recipe made by Juha Reunanen. We checked the combined cubes for faint star-like persistence effects (see van Starkenburg et al. \cite{prep1}) but there were none.

We measured the position and FWHM of the PSF stars at the wavelength range (averaged over 40 pixels) of the H$\alpha$ and [\ion{N}{ii}]$\lambda$6583 line. We used these positions to combine the data in pairs (``second subtraction'') for reasons described in Appendix \ref{app}. The last two science frames were observed at the end of the night. There was no time left to observe PSF stars for these frames. We used the positions of the previous observation block (OB) for this pair (which will get very low weight in the final combination anyway because the seeing was poor).

Then, we flux calibrated the data using the PSF and standard star data. The flux of the PSF stars correlates well with $1/\textrm{airmass}$, and we used this relation to do a relative flux calibration for all pairs of science frames. We also observed four bright standard stars during the night, but those did not show this tight correlation between airmass and flux (after correcting for different integration times and luminosity). Two standard stars gave consistent results (with the PSF star relation), the third differed in the amount of atmospheric absorption, the fourth was more noisy. Fortunately, H$\alpha$ is not at a relevant wavelength for atmospheric absorption. We choose to use the average of the two consistent standard stars for absolute flux calibration and ignored the other two standard stars.

We used the FWHMs of the PSF stars to weigh the data when combining all data. We found the best result in terms of resolution and signal-to-noise ($S/N$) when weighting with the square of the FWHM in the y direction (which is unaffected by the complex FOV reconstruction). We measured the seeing in the combined image from the combined PSF star image (using the same weights, at wavelength of H$\alpha$ and [\ion{N}{ii}]$\lambda$6583) and found 0\farcs55. The FWHM in the x direction is 20\% larger. 

Further details on the reconstruction of the FOV using the PSF stars are provided in Appendix \ref{app}.

\section{Results}\label{res}

\begin{figure}[t]
  \resizebox{\hsize}{!}{\includegraphics{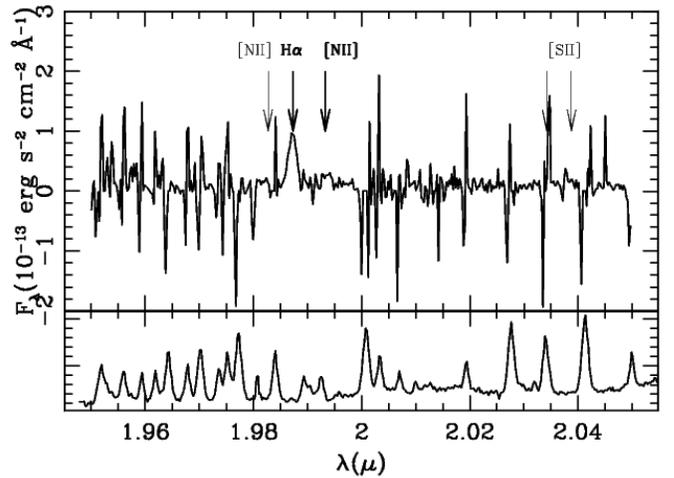}}
  \caption{The integrated 1D spectrum of F257. We detect H$\alpha$ and [\ion{N}{ii}]$\lambda$6583. The positions of the [\ion{N}{ii}]$\lambda$6548 and [\ion{S}{ii}]$\lambda$6718/6733 are also marked in grey.  The logarithm of the sky spectrum with OH lines is shown in the bottom panel. }
  \label{1dspec}
\end{figure}

The final integrated 1D spectrum of F257 is shown in Fig. \ref{1dspec} and shows H$\alpha$ at redshift 2.028 together with [\ion{N}{ii}]$\lambda$6583 ([\ion{N}{ii}]$\lambda$6548 falls on top of a bright OH line). We do not detect the [\ion{S}{ii}] lines. The continuum is very faint but clearly detected in the median collapsed cube (excluding the wavelength range of the emission lines). The rest frame equivalent width (EW) of H$\alpha$ is $\sim160 \AA$ where the continuum flux was calculated from a linear interpolation of the observed H and K magnitudes from the FIRES data. 

The [\ion{N}{ii}]/H$\alpha$ ratio is $\sim0.22$ (averaged over the whole galaxy) which is consistent with ionization by star formation. The metallicity calculated from this ratio is $12+\log(O/H) = 8.6$  which corresponds to slightly subsolar metallicity (Denicolo et al. \cite{denicolo}, solar metallicity is $ 12+\log(O/H) = 8.91$). The [\ion{N}{ii}]/H$\alpha$ ratio is slightly larger in the centre of the galaxy ($[\ion{N}{ii}]/H\alpha  \sim 0.4$) which gives metallicity $ 12+\log(O/H) = 8.8$.  The [\ion{N}{ii}]/H$\alpha$ ratio is also consistent with ionization by star formation in the centre of the galaxy. Ionization by an AGN would would give larger a larger [\ion{N}{ii}]/H$\alpha$ ratio.  

Using the Kennicutt (\cite{kennicutt}) relation for a Salpeter IMF, the H$\alpha$ flux corresponds to a star formation rate (SFR) of $116 \textrm{ M}_\odot \textrm{yr}^{-1}$  uncorrected for extinction. Using $A_V=1.0$ from Wuyts et al. (\cite{wuyts}), the SFR becomes $232 \textrm{ M}_\odot \textrm{yr}^{-1}$. This value is very close to the value found when dividing the best fit mass and age from the SED fitting: $ 219 \textrm{ M}_\odot \textrm{yr}^{-1} $.

\begin{figure}
  \resizebox{\hsize}{!}{\includegraphics{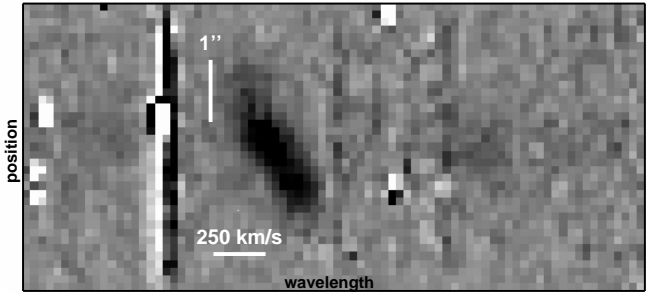}}
  \caption{Synthetic long slit spectrum. Scales are indicated in the figure. Note the presence of faint [\ion{N}{ii}] emission. }
  \label{longslit}
\end{figure}

\begin{figure}
  \resizebox{\hsize}{!}{\includegraphics{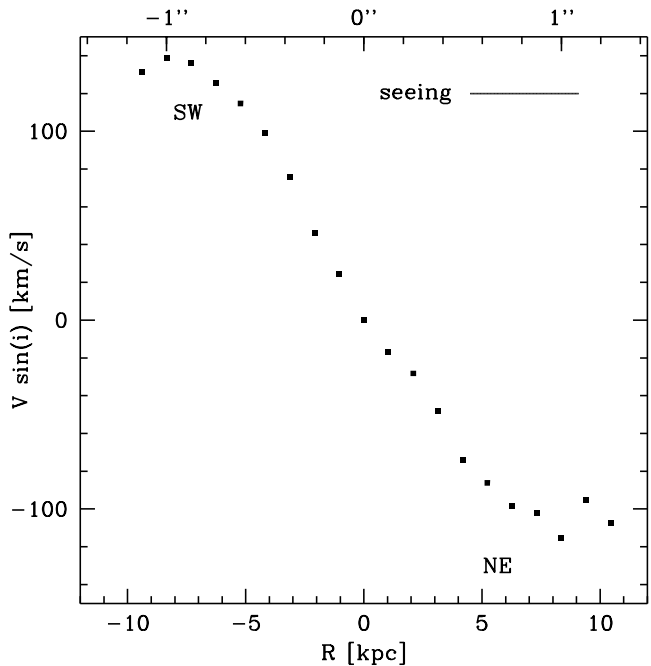}}
  \caption{Rotation curve as measured from the synthetic long slit spectrum by fitting a gaussian to each spatial position in Fig. \ref{longslit}. The formal fit errors are comparable to the size of the symbols, except for an increase in the outer parts. The points are correlated due to the 0\farcs55 seeing.  }
  \label{rclongslit}
\end{figure}

The FWHM of the H$\alpha$ line is  $244 \textrm{ km s}^{-1}$ after correction for instrumental resolution. Again, there is no evidence for the presence of an AGN in the form of broad lines. Stepping through the wavelength slices of the cube, one notices immediately the wavelength shift of H$\alpha$. A synthetic long slit spectrum (along the major axis, position angle $32 \degr$ counterclockwise from the north) made from the cube integrated over the minor axis of the galaxy with slit width 1\farcs25 shows the tilted emission line that flattens at at least one side, see Fig. \ref{longslit}. The RC measured from the synthetic long slit spectrum is shown in Fig. \ref{rclongslit}. We will discuss the full velocity field analysis in the next section.

\begin{figure*}
\includegraphics[width=9cm]{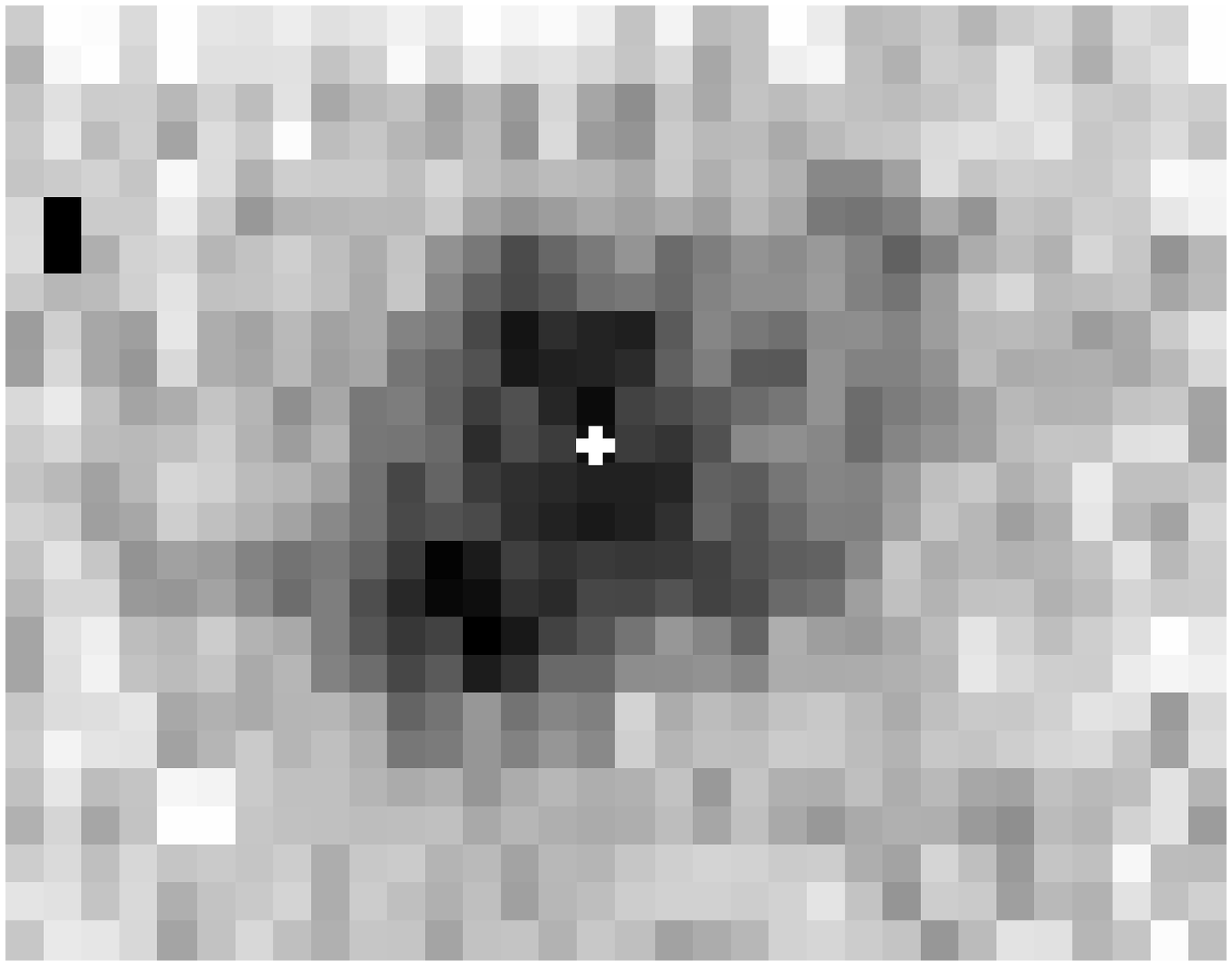}
\includegraphics[width=9cm]{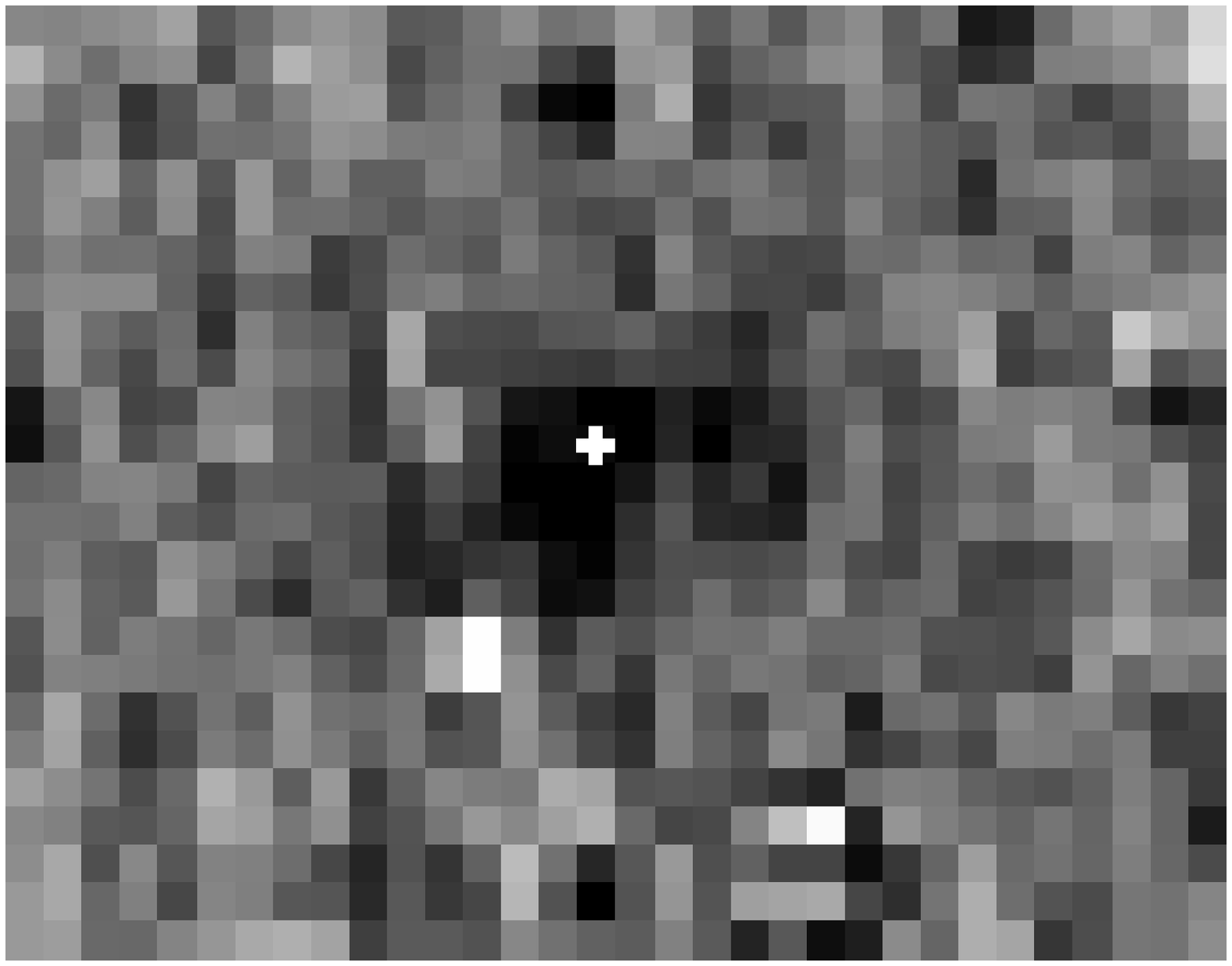}
  \caption{a. H$\alpha$ line map. b. [\ion{N}{ii}]$\lambda$6583 line map. Image size is again 3\arcsec~ times 4\arcsec~ and the images have been rotated so that east is to the top and north is to the right. The white crosses mark again the (K band) center (here determined from the PSF star positions).}
  \label{linemap}
\end{figure*}

From our final data cube, we extracted line maps of H$\alpha$ and [\ion{N}{ii}]$\lambda$6583, shown in Fig. \ref{linemap}a and  \ref{linemap}b. The H$\alpha$ line map shows clumps which resemble the clumps of the F814W image. This is not surprising as both F814W(=rest frame UV) and H$\alpha$ trace star formation.

The [\ion{N}{ii}]$\lambda$6583 line map shows a much more concentrated distributions, although [\ion{N}{ii}]$\lambda$6583 is detected over the whole galaxy. If the inferred metallicity gradient is also present in the stellar population, this would put the highest metallicity and hence oldest stars in the centre of the galaxy, consistent with the redder color in the centre of the galaxy as found by Labb\'e et al. (\cite{labbe}). The [\ion{N}{ii}]$\lambda$6583/H$\alpha$ line ratio (0.2-0.4) is consistent with star formation over the whole galaxy.

\section{Velocity field}

We fitted a Gaussian to the H$\alpha$ line at every position in the cube. We did this by hand using IRAF's splot because the $S/N$ ratio of individual spectra is low and nearby OH line residuals confuse automatic fitting routines. For the outer parts of the galaxy, we were able to get a few additional data points by adding the spectra of 4 pixels (which were used as one pixel in the modeling). The resulting velocity field is shown in Fig. \ref{velfields}a. It shows a remarkably regular spider diagram of a rotating disk.

\subsection{Tilted ring model}\label{tiltedring}

\begin{figure*}
   \centering
\includegraphics[width=9cm]{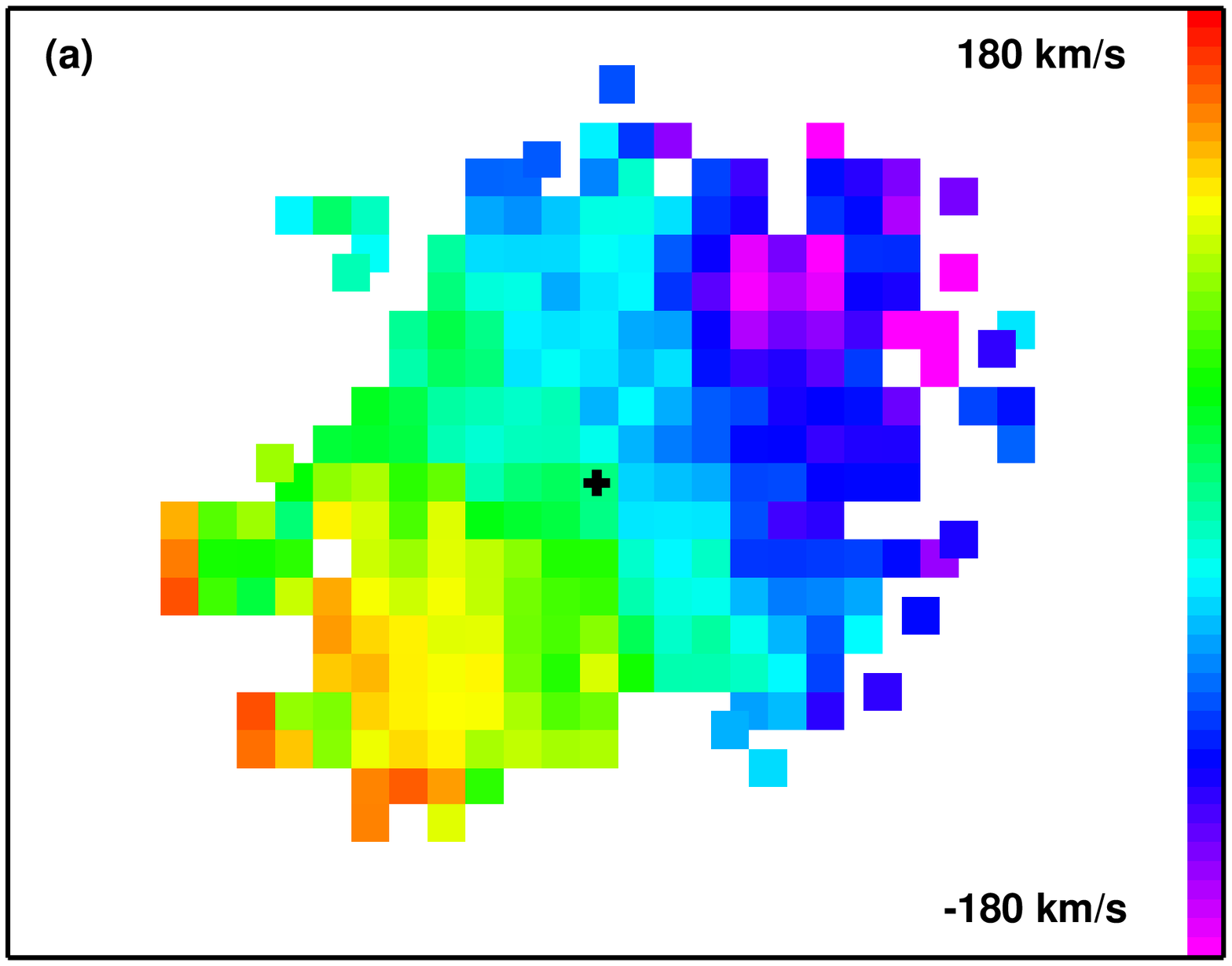}
\includegraphics[width=9cm]{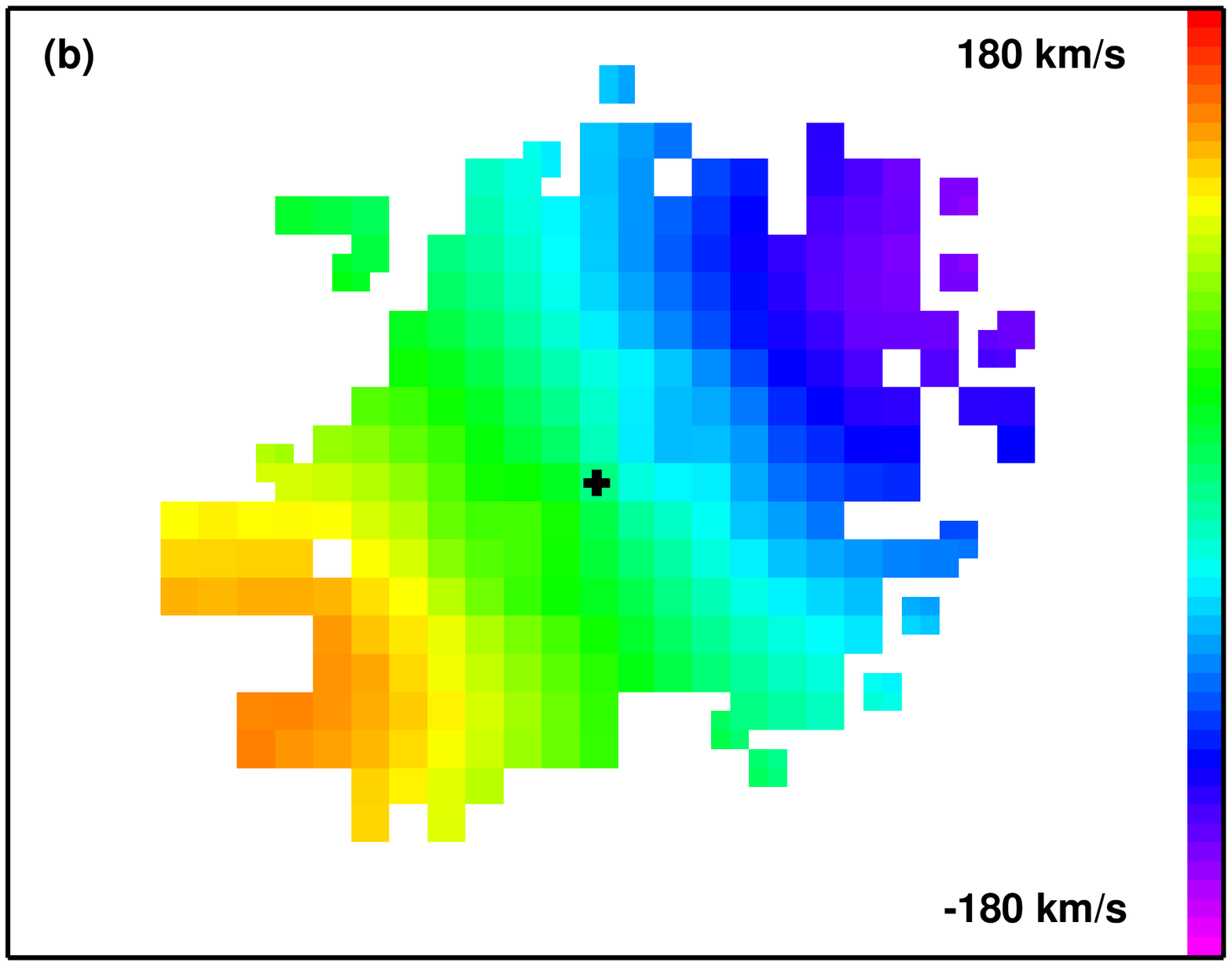}
\includegraphics[width=9cm]{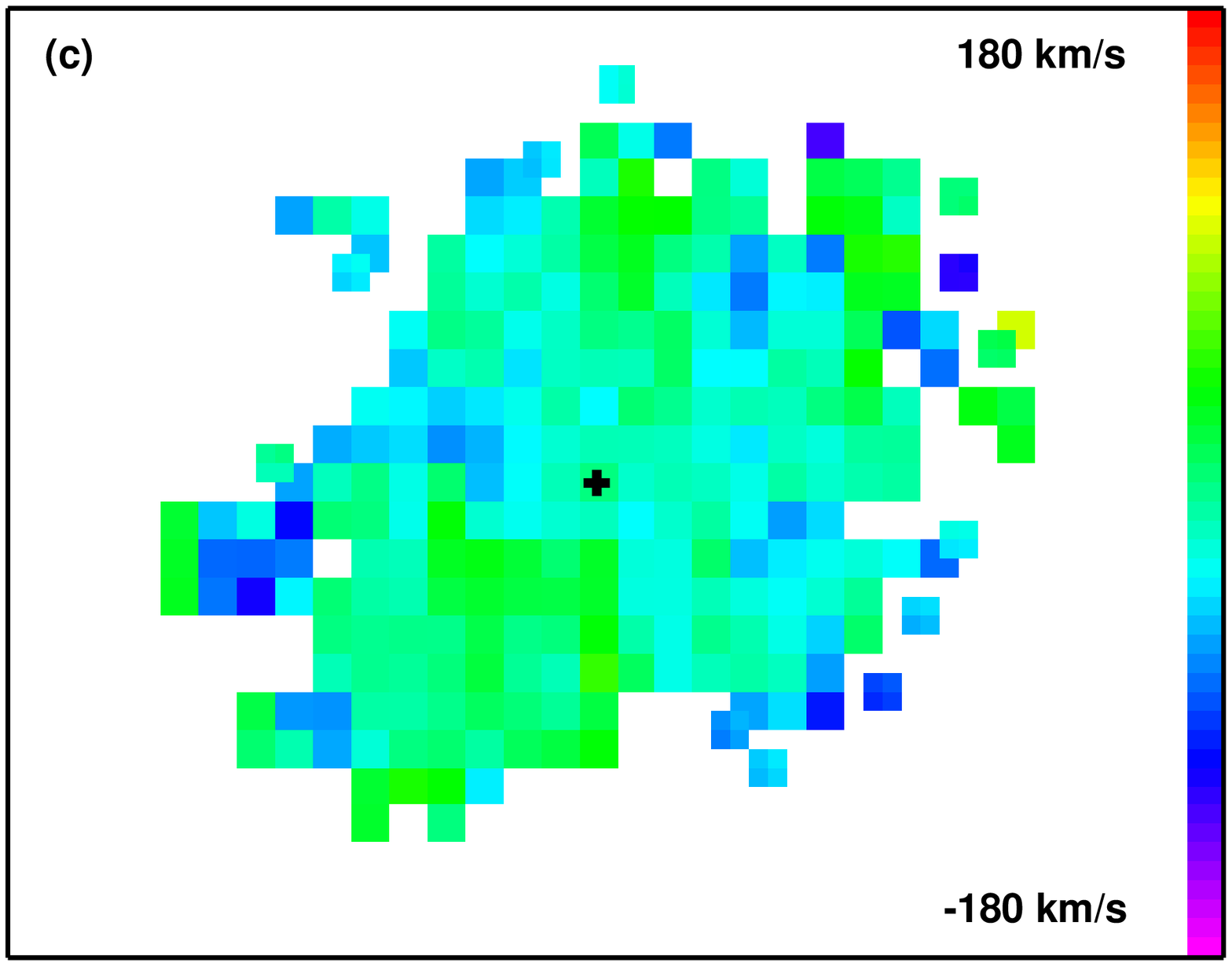}
\includegraphics[width=9cm]{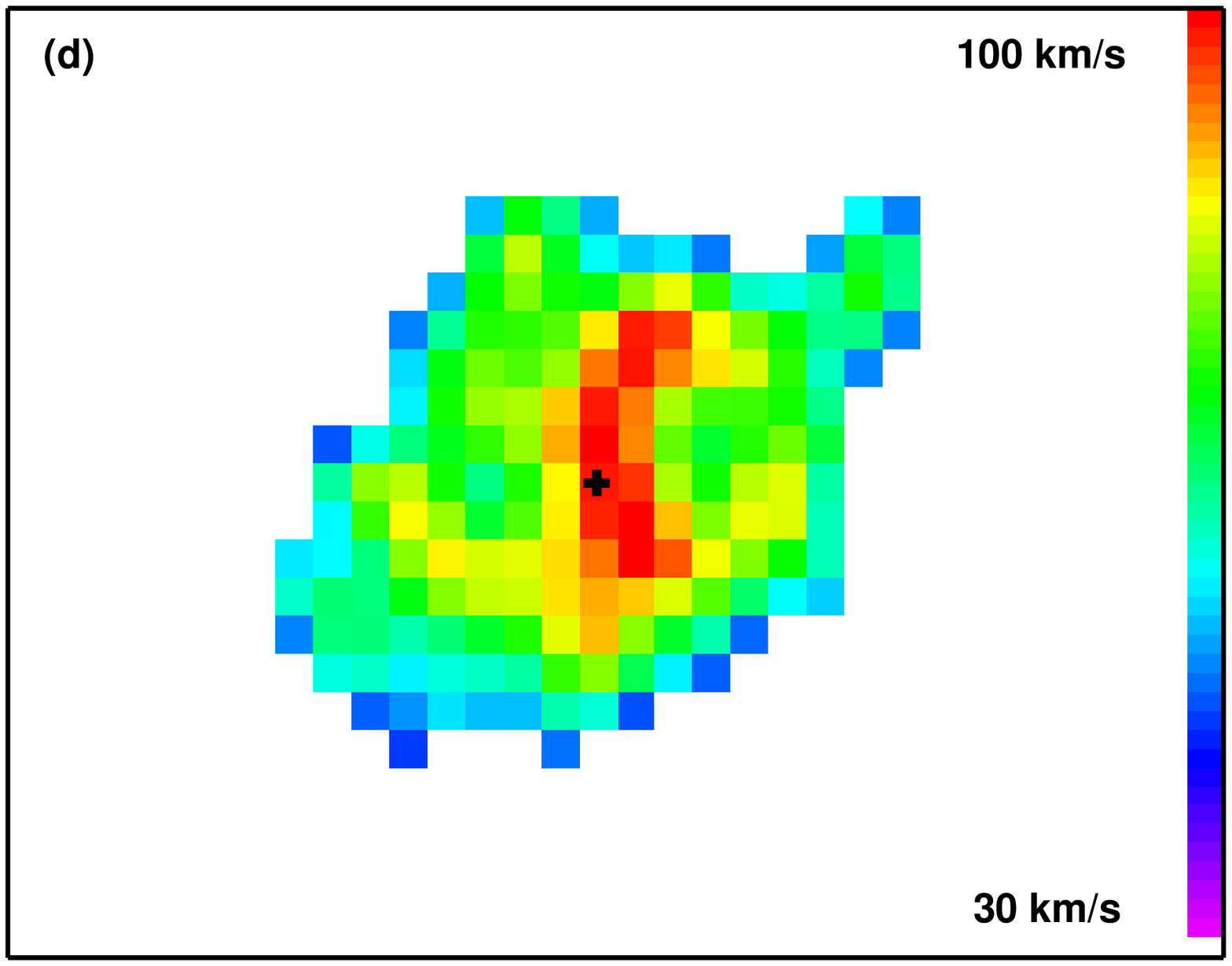}

   \caption{a. The observed velocity field (not corrected for inclination). Some points at the edges are the average of four pixels. b. Best fit model velocity field. c. Residuals from the fit. d. Velocity dispersion field. The extent is smaller than for the velocity field due to the lower $S/N$. Instrumental resolution is $32 \textrm{ km s}^{-1}$. Note that the velocity scale is different for this figure. The spatial scale and orientation is the same as in previous figures and the cross point marks again the (dynamical and photometric) center. }
       \label{velfields}
\end{figure*}

\begin{figure}
  \resizebox{\hsize}{!}{\includegraphics{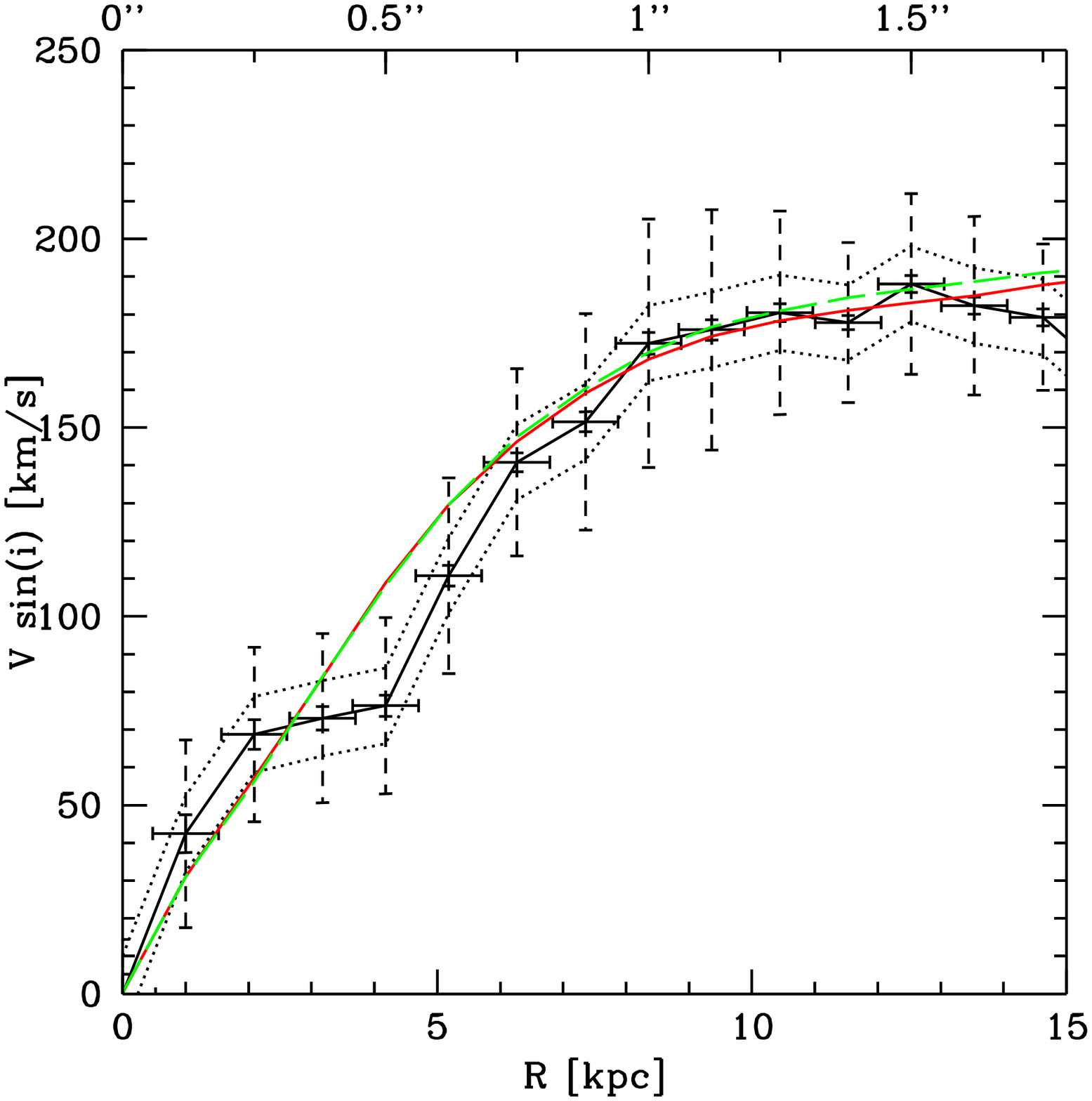}}
  \caption{Rotation curve from the velocity field model shown in Fig. \ref{velfields}b. To guide the eye, the data points (with errorbars) are connected by the (black) thick line and the dotted lines are $+10 \textrm{ km s}^{-1}$ and $-10 \textrm{ km s}^{-1}$. The horizontal errorbars indicate the width of the rings. The full vertical errorbars are the errors on the fitted velocity, the dashed vertical errorbars are the range of velocities in the ring. The full (red) line is the best fit RC as discussed in Section \ref{psfrc} ($a=1.0$, $b=0.5$ and  $i=54\degr$). The long dashed (green) line shows an almost identical fit ($a=1.1$, $b=0.5$ and $i=64\degr$).  }
  \label{rcvelfield}
\end{figure}

We continued by fitting a simple model to the data using GIPSY (van der Hulst et al. \cite{hulst}, Vogelaar \& Terlouw \cite{vogelaar}), which contains a software package designed for fitting tilted rings to HI velocity fields. Free parameters for each ring are  in principle  the centre of the galaxy, its inclination, the position angle of the major axis, systemic velocity, expansion velocity and the rotation velocity. For simplicity, we fixed the centre to the centre of the FIRES data, and the inclination and position angle of the major axis to those of the FIRES  K band image. We also fixed the systemic velocity to the velocity of the central pixel, and fixed the expansion velocity to zero. This leaves only the rotation velocity as a free parameter for each ring. The resulting model is shown in Fig. \ref{velfields}b and the RC in Fig. \ref{rcvelfield}.  We will discuss the properties of the RC and the parameters of the the fit in the next section.

\subsection{Residuals from fit}
In Fig. \ref{velfields}c we show the residuals of our tilted ring model fit. The lack of  systematic residuals is a strong indication for a rotating disk. The mean  absolute residual is $27 \textrm{ km s}^{-1}$ while large residuals ($> 52 \textrm{ km s}^{-1}$) only occur at the noisy outer edges of the galaxy. The mean residual excluding these points is $21 \textrm{ km s}^{-1}$.

These residuals are slightly larger than  those of Wright et al. (\cite{wright}) who observed a $z\sim1.5$ disk galaxy and found a mean residual of $13 \textrm{ km s}^{-1}$. They note that this is only slightly larger than the deviations observed for local spirals ($\sim 10 \textrm{ km s}^{-1}$). They explain  their larger deviations by beam smearing. Our seeing limited observation suffer even more from beam smearing than their AO observations.  Modeling  our residuals while taking into account  these effects would require assumptions on the intrinsic distribution of the H$\alpha$ flux and dispersion and the intrinsic velocity field.  We did not attempt to make such a model as the many uncertainties would not add much to the current conclusion: the small random residuals provide evidence that F257 is a disk galaxy with a regular velocity field quite similar to that of local galaxies.

These results are in contrast to many other kinematic measurements of high-$z$ galaxies which show significant deviations from regular rotation, e.g. the $z=2.38$ galaxy for which Genzel et al. (\cite{genzel}) made a rotating disk model shows offsets up to $\sim 170\textrm{ km s}^{-1}$ (comparable to the maximum rotation curve velocity) and deviations from a simple disk geometry. 

\subsection{Dispersion field}
A peak in the measured velocity dispersion at the centre of rotation would be consistent with a rotating disk model. Although the $S/N$ of individual spectra is poor, we made an attempt to measure the velocity dispersion field. We measured the dispersion on the raw 1D spectra, corrected for instrumental resolution and then smoothed the dispersion map with a 2 pixel width Gaussian to increase $S/N$. The dispersion map is shown in Fig \ref{velfields}d. 
The dispersion is larger in the centre (maximum is $103 \textrm{ km s}^{-1}$) and smaller in the outer parts ($ \sim65 \textrm{ km s}^{-1}$). The shape of the dispersion peak appears elongated, but the $S/N$ is insufficient to make any statement about the precise shape.

Peaks in the centre are not commonly observed at this redshift: the $z=2.4$ galaxy of Genzel et al. (\cite{genzel}) does show a peak in the centre but the overall shape is distorted, F\"orster-Schreiber et al. (\cite{forster}) find a peak for 1 out of 6 galaxies. Even at $z=0.6$, Flores et al. (\cite{flores}) find only 35\% rotating disks (in their definition: the major axis of the velocity field coincides with the major axis of photometry plus a peak in centre of the dispersion field) in a representative sample of emission line galaxies.

\subsection{Clumps}
The H$\alpha$ line map and F814W image showed a clumpy structure. This raises the question to what extent the observed velocity field is biased by the inhomogeneous surface brightness distribution of H$\alpha$, emphasizing selected velocity ranges.
The flux in clumps is 10-30 \% of the total flux (measured in F814W), so although prominent in the image, they will not dominate the velocity field observed. We checked this in the model velocity fields where the F814W image was used as model for the true H$\alpha$ surface brightness distribution (see Sect. \ref{psfrc}): there is no visible distortion of the velocity field at the location of the clumps.
In addition, the smooth appearance of the velocity field and the agreement of kinematic and photometric (K band) major axis argues that the velocity field is dominated by the stellar mass.
We conclude that the clumpy structure of the H$\alpha$ surface brightness distribution has no significant impact on the velocity field observed.

\subsection{Summary}
The velocity field of F257 reveals a remarkably regular spider diagram. A tilted ring model, leaving only the rotation velocity as a free parameter for each tilted ring, fits the observations very well. The residuals from the fit are random and  much smaller than in many other high-$z$ galaxies with similar kinematic data (e.g. Genzel et al. \cite{genzel}). The dispersion field shows an extended peak in the centre. The clumpy distribution of star formation activity has no significant impact on the velocity field observed. In the next section, we will discuss the RC found including the effect of PSF smearing  on $V_{flat}$ and $R_{flat}$, which we have ignored so far.

\section{Rotation curve}

\begin{figure}
  \resizebox{\hsize}{!}{\includegraphics{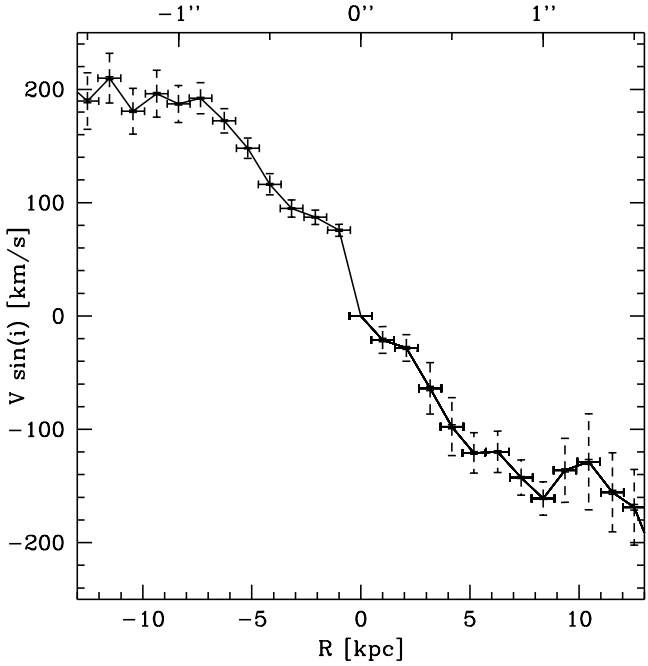}}
  \caption{Rotation curve from the velocity field model, approaching and receding side fitted separately. To guide the eye, the data points are connected by the thick line. The errorbars have the same meaning as in Fig. \ref{rcvelfield}.
 }
  \label{rc2}
\end{figure}

The RC from the tilted ring model for the velocity field as discussed in the previous section was shown in Fig. \ref{rcvelfield}. It rises until it reaches its maximum velocity of $\sim 170 \textrm{ km s}^{-1}$ (not corrected for inclination) at $\sim 1 \arcsec$ and then stays flat. Note that this velocity is significantly larger than the velocity found in the synthetic long slit spectrum which is probably due to the integration over the width of the synthetic slit. 
The synthetic long slit spectrum (see section \ref{res} and Fig. \ref{rclongslit}) suggested that the RC flattens at both the receding and approaching ends at almost the same velocity. We repeated the fit of the velocity field as done in Sect. \ref{tiltedring}, but now fitting the approaching and receding sides seperately. The resulting RC, shown in Fig. \ref{rc2}, confirms that the RC flattens at the receding and approaching half at the same radius and velocity. Altough the H$\alpha$ flux distribution is not symmetric (see linemap in Fig. \ref{linemap}), this appears to have no significant influence on the best fit RC. 
For simplicity and $S/N$, we use the fit of the entire velocity field in the remainder of this paper.

The errorbars on the RC were determined as follows: GIPSY calculates for each ring the error on the velocity $\Delta V$ and the range of velocities $\sigma_V$. The first is $\sim 3 \textrm{ km s}^{-1}$, the latter on average $26 \textrm{ km s}^{-1}$. Both $\Delta V$ and $\sigma_V$ are shown in Fig. \ref{rcvelfield}.
Here, equal weights are assigned to all pixels. Within a single ring, all pixels have approximately the same H$\alpha$ flux and hence the same errorbar. The $S/N$ drops from the centre outwards while at the same time the number of pixels in the fit increases. Assuming the errorbar on the velocity of a single pixel scales with the inverse of the square root of the H$\alpha$ flux, the errorbars on the RC are approximately constant except for the inner few rings and for $r \gtrsim 1\farcs4$ where the errorbars increase.

\subsection{Parameters of the fit}

We now turn to the parameters we fixed when we made a model of the velocity field: the position angle of the major axis, inclination, centre, systemic velocity and expansion velocity.  This is done in order to investigate if their values can be reproduced from the velocity field alone, and if so, if they are consistent with the values we assumed in Section \ref{tiltedring}. We also want to know the effect of small variations of these parameters on the best fit RC to get an estimate of the errorbars of the RC.  It is not possible to leave all these parameters as free parameters at the same time. The number of free parameters is then too large for the number of independent data points, and the resulting best fit parameters for each ring vary strongly from ring to ring in an unphysical way (e.g. the inclination varies between 10 and 80\degr). Therefore, we study only the case of two free parameters: the rotation velocity plus one other parameter. 

\begin{enumerate}
  \item If the angle of the major axis is left as the second free parameter, one finds a consistent result between the orientation of the rings and the K band isophotes. 
This indicates that the orientation of the stellar mass distribution as traced by the K band light is consistent with the gravitational field as traced by the velocity field. 
Varying the position angle caused negligible changes in the RC of the fit. 
 
 \item When the expansion velocity is the second free parameter, we find $V_{exp}\sim 0 \textrm{ km s}^{-1}$ with small scatter in all rings.

  \item Leaving the x and/or y position of the centre as free parameter(s), we find small (1-2 pixel) offsets from the K band centre, varying from ring to ring and with no significant effect on the RC found. When we vary the fixed centre around the K band centre, we also do not find systematic changes in the RC found. We therefore keep the centre at the K band centre. 
 
  \item We were not able to determine the inclination from the velocity field. If the inclination is a free parameter, the best fit inclination for each ring does not follow an ordered pattern, indicating that the $S/N$ ratio does not allow determination of the inclination from the velocity field alone. However, fixing the inclination to different values causes small but systematic offsets in the flattening velocity: more edge-on galaxies have slightly smaller flattening velocities than more face-on galaxies. The difference between $i=64\degr$ and $i=44\degr$ is $\sim15 \textrm{ km s}^{-1}$. Without the K band constraint on the inclination, this small systematic difference would result in a larger uncertainty in $V_{flat}$.

  \item Leaving the systemic velocity as a free parameter, we find a small systematic offset from the velocity of the central pixel which we explain by the small difference in $V_{flat}$ of the receding and approaching sides of $\sim 10 \textrm{ km s}^{-1}$. The offset is smaller than the differences in $V_{flat}$ we find fitting with different parameters (see conclusion below). Therefore, we ignore it in the remaining analysis. 

  \item GIPSY also offers different weighting possibilities to the pixels within a ring but this also does not significantly change the fitted RC.

\end{enumerate}

We were able to reproduce all parameters except the inclination from the velocity field alone. They all agree within the errorbars with the a priori assumend values. The agreement between the centre and position angle of F257 in the Ks band image and the velocity field, and the agreement between the systemic velocity and the velocity of the centre, both strongly support the rotating disk interpretation. 

Changing the parameters in the fit except the inclination results in random changes of the RC of the simplest model. The error on $V_{flat}\sin(i)$  derived by studying the variations due to the free parameter choice of the tilted ring model is $\sim 15 \textrm{ km s}^{-1}$. Varying the inclinations causes systematic offsets, that will become important when converting the observed velocities to true inclination corrected velocities. Therefore, we will focus on the inclination hereafter and keep all other parameters fixed.

\subsection{PSF}\label{psfrc}

So far, we have ignored smearing of our velocity field by seeing. Although F257 is a large galaxy (we measure H$\alpha$ over $\sim3\arcsec$ diameter), the seeing of 0\farcs55 has smeared the velocity field observed significantly. We will now model the effect of the PSF on the RC observed. For simplicity, we will focus on the RC only and not on the full velocity field. Specifically, we would like to know what the true $V_{flat} \sin(i)$ is, whether we can extract more information about the inclination of the galaxy and the radius where the RC flattens, $R_{flat}$.

We generated our model velocity fields by taking the observed RC and modify it according to
\begin{equation}V_{model}(r) = \left\{ \begin{array}{rl} 
  a * V_{obs}(br) & \textrm{if }  br < r_{max}  \\
  a * V(r_{max}) & \textrm{if } br > r_{max} \ 
\end{array} \right.
\end{equation} 
where $V_{obs}(r)$ is the velocity field observed at distance $r$ from the centre, $r_{max}$ is the largest radius for we could measure the rotation velocity, $V_{model}$ is the velocity of the model, constant $a$ scales the maximum velocity of the model and constant $b$ scales the radius where the RC flattens. 
We then made synthetic model velocity fields (VF) for different $a$ and $b$ and created synthetic observations using
\begin{equation}  VF_{model,obs} = \frac{( VF_{model,a,b} \times F_{true}) \otimes PSF }{  F_{true} \otimes PSF }\end{equation}
where $VF_{model,obs}$ is the model velocity field as ``observed'', $VF_{model,a,b}$ is the intrinsic model velocity field for constants $a$ and $b$ and $F_{true}$ is the true flux distribution of the emission line. $F_{true}$ is of course also unknown. We used the F814W image, assuming that H$\alpha$ and F814W (=rest frame UV) have the same distribution (as is likely, see section \ref{res}). We used the elliptical gaussian as reconstructed from the PSF star images for the PSF. We then fitted a RC to each model for different inclinations.

We found fitting RCs for $a=1-1.1$. Larger values of $a$ overpredicted $V_{flat}\sin(i)$ or $R_{flat}$ (depending on the inclination), smaller values could not reproduce the maximum velocity observed. Our best fit $V_{flat}\sin(i)$ is then $179 \pm 9 \textrm{ km s}^{-1}$.

\begin{figure}
  \resizebox{\hsize}{!}{\includegraphics{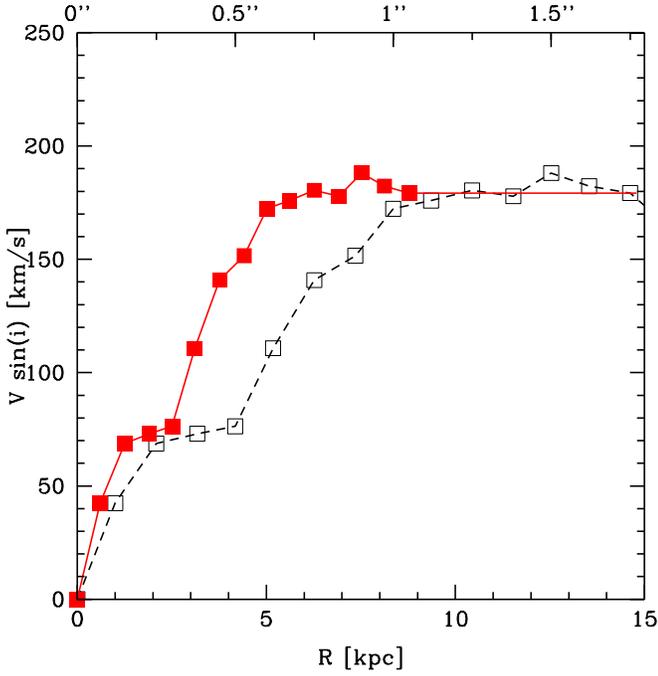}}
  \caption{The observed RC (open squares connected by the (black) dashed line) and the PSF corrected RC (full squares connected by the full (red) line). }
  \label{rcpsf}
\end{figure}

The best fit RCs have $b=0.5-0.6$, although acceptable fits can also be found for slightly larger values up to $b=0.8$. If $R_{flat,obs} \approx 1\arcsec$, $R_{flat,true}$ becomes $0.55 \pm 0.15 \arcsec$ or $ 4.6 \pm 1.3 \textrm{ kpc}$. Beam smearing has thus a profound effect on the observed flattening radius of the RC. This is illustrated in Fig. \ref{rcpsf}, where we show the observed RC and the PSF corrected RC. Altough $R_{flat,true}$ is of the order of the seeing, the radius where the RC flattens is resolved: in the velocity field, the (true) flattening radius of the approaching and receding half are $2 R_{flat,true}$ apart. 

Including the effect of beam smearing in our analysis, we  were not able to constrain the inclination better as is shown in the almost identical fits with different inclination in  Fig. \ref{rcvelfield}.

To summarise, we find $V_{flat} \sin(i) =  179\pm17 \textrm{ km s}^{-1}$, 
$R_{flat} = 4.6 \pm 1.3 \textrm{ kpc}$ and no better constraints on inclination after correction for beam smearing. 
Using our best estimate of the inclination (from the K band photometry) $i = 54 \pm 5\,^{\circ} $, we find $V_{flat} = 221 \pm 22 \textrm{ km s}^{-1} $.

\subsection{RC shape}
The RC of F257 flattens at $ 4.6 \pm 1.3 \textrm{ kpc}$. 
In the local universe, the $R_{flat}$ depends on the bulge-to-disk ratio, larger bulge-to-disk ratio's correlating with steeper rising RCs (Noordermeer \cite{noordermeer}). The RCs of local early-type spiral galaxies rise to their maximum velocity in $< 1 \textrm{ kpc}$ (Noordermeer \cite{noordermeer}), which is clearly not the case for F257 as expected from the pure exponential disk profile found by Labb\'e et al. (\cite{labbe}). Swaters, Madore \& Trewhella  (\cite{swaters}) compare the RCs of their low surface brightness late-type galaxies with three late-type high surface brightness galaxies from Begeman (\cite{begeman}). They all flatten at $2-5 \textrm{ kpc}$. So, the shape of the RC of F257 appears consistent with late type local spirals.

Slowly rising RC at high redshift have also been noted by Swinbank et al. (\cite{swinbank}). They studied gravitationally lensed galaxies at $z\sim1$ using integral field spectroscopy  and found that the galaxies with stable disk kinematics have slowly rising rotation curves and suggest that these have lower bulge-to-disk mass ratio's than their local counterparts.

\section{Masses}\label{masses}

From our model for the velocity field we calculate the total dynamical mass of the galaxy within $r=1\farcs5$
\begin{equation} M_{dyn} = \frac{v^2 r}{G}  \end{equation}
and find $M_{dyn} = 1.4 \pm 0.3 \times 10^{11} \textrm{ M}_\odot$ (depending on geometric assumptions, this equation is found with different constants of order 1; when we compared our finding with those of others, we used their constant.)

Wuyts et al. (\cite{wuyts}) made an extensive study of stellar masses for high redshift galaxies in HDF-S and found for F257 $3.5^{ +0.5}_{ -0.3} \times 10^{10} \textrm{ M}_\odot $ (using a Salpeter IMF). Using a 'diet' Salpeter IMF, a Salpeter IMF with 30\% less mass to correct for the overprediction of low mass stars (as for example Bell \& de Jong \cite{bell}), would reduce this mass to $2.5 \times 10^{10} \textrm{ M}_\odot $.

A crude estimate of the gas mass may be found using the global Schmidt law for star forming galaxies which relates SFR and gas surface density (Kennicutt \cite{kennicutt}, see also Bouch\'e et al. \cite{bouche} for evidence that this relation also holds at high redshift). Using the SFR calculated from the H$\alpha$ flux and the total area over which we observe H$\alpha$, this gives a total gas mass of  $ 5.3 \times 10^{10} M_{\odot}$, and after correcting for extinction ($A_V=1.0$, Wuyts et al. \cite{wuyts}) $8.7 \times 10^{10} M_{\odot}$. Although the interpretation of this number is uncertain due to the large scatter in the Kennicutt relation of  $\pm 0.3$ dex, it is comparable to or larger than the stellar mass, which is consistent with picture that this galaxy is building up its stellar mass at a fast rate.

From these masses, we get the following mass ratio's (Salpeter IMF, extinction corrected): \\
- $M_{gas}/M_* =  2.5 $  \\
- $M_*/M_{dyn} = 0.25  $ ; \\
- $M_{baryon}/M_{dyn} = (M_* + M_{gas})/ M_{dyn} = 0.9 $ \\
Local galaxies with similar gas-to-stellar mass ratio are rare, and such galaxies have stellar masses $\sim 1 - 2$ orders of magnitude smaller than the stellar mass of F257 (Kannappan \cite{kannappan}).  Similar gas-to-dynamical mass ratio's are locally only found in extreme late type galaxies (Binney \& Merrifield \cite{binney}). 
The gas-to-stellar mass ratio is consistent with the $z\sim2$ rest frame UV selected galaxies of Erb et al. (\cite{erb06}). The stellar-to-dynamical mass ratio is at the low end of the range found by Erb. et al. They find galaxies with lower stellar-to-dynamical mass ratio's to be younger, and F257 is indeed a relatively young galaxy as found by Wuyts et al. Although F257 is photometrically young, its dynamical mass is at the high end of the range found by Erb et al.

The sum of stellar and baryonic mass is $1.2 \times 10^{11} M_\odot$ which is almost the same as the dynamical mass. Within the errorbars, there is no evidence for dark matter within $ r < 12 \textrm{ kpc}$ in this galaxy.

F257 appears to be kinematically like a local late type spiral galaxy in many respects. We have seen that its velocity field is remarkably regular, a tilted model leaves small residuals and that the dispersion increases towards the centre. The RC slowly rises and then remains flat like those of local late type spirals. F257 is a very gas rich spiral compared to local galaxies.  We will now further investigate the nature of this disk galaxy in terms of dynamical support and specific angular momentum. We will then investigate if it follows the local B, K, stellar and baryonic mass TFR.

\section{Rotational support and specific angular momentum}

Galaxies can be supported by rotation and/or dispersion. Disk galaxies are supported by rotation. $V/\sigma$ is a measure of the dynamical support of a galaxy. There is no general agreement on what to use for $\sigma$ for disk galaxies. Puech et al. (\cite{puech}) use as $\sigma$ the velocity dispersion in the $z$ direction which they calculate from the mean observed $\sigma$ in the outer regions of their disks. We followed Puech et al. (\cite{puech}) and found an average $\sigma_{z}$ of $57 \textrm{ km s}^{-1}$ and hence $V/\sigma_{z} = 3.9 \pm 0.4$. This is comparable to the median  $V/\sigma_{z}$ found by Puech et al., $V/\sigma_{z} = 3.8 \pm 2$, for their $z=0.6$ rotating disk galaxies. Their local comparison sample has median $V/\sigma_{z} = 6.1 \pm 1.1$ while their  $z=0.6$ perturbed rotating disks have $V/\sigma_{z} = 2.4 \pm 2.5$. Although the gas disk of F257 is heated compared to local galaxies, it is dynamically mainly supported by rotation, comparable to the $z\sim0.6$ galaxies of Puech et al. (\cite{puech}).

An important question in galaxy formation theory is when disk galaxies acquired their angular momentum. The specific angular momentum of a thin rotating disk is 
\begin{equation} \gamma = 2R_dV_{max} \label{eq}\end{equation}
which gives $\log(\frac{\gamma}{\textrm{kpc km s}^{-1}}) = 3.2 \pm 0.2 $ for F257. This is slightly smaller than the average specific angular momentum of galaxies of the same mass at redshift zero, but consistent with the specific angular momentum - velocity relation for local galaxies (Puech et al. \cite{puech}, Navarro \& Steinmetz \cite{navarros}). Salucci et al. (\cite{salucci}) also found their $z\sim1$ galaxies to be consistent with the local specific angular momentum - velocity relation.  For other mass distributions, the factor 2 in equation \ref{eq} is slightly different, but this does not change our conclusion.

Another measure of angular momentum is the spin parameter $\lambda$ . We followed again Puech et al. (\cite{puech}) using their expression for $\lambda$ in terms of disk scale length and flat rotation velocity and found $\lambda=0.019\pm0.006$ which is comparable to the values found by Puech et al. (\cite{puech}) for their $z\sim0.6$ rotating disk sample and their local reference sample and also close to the estimate for the local value made by Tonini et al. (\cite{tonini}) ($\lambda \sim 0.03$).

We conclude that F257 is a dynamically fully formed disk galaxy: it is mainly rotationally supported and  has the same specific angular momentum as local late type spirals. It is important to note that this makes F257 excellently suitable for a comparison with the local TFRs, which we will do now.

\section{Tully-Fisher relations}
Having established that F257 is dynamically very much like a local disk galaxy, we compare it to local TFRs to compare its baryonic components to dynamically similar local galaxies.

\subsection{The rest frame B and K band TFR}
\begin{figure*}
   \centering
\includegraphics[width=9cm]{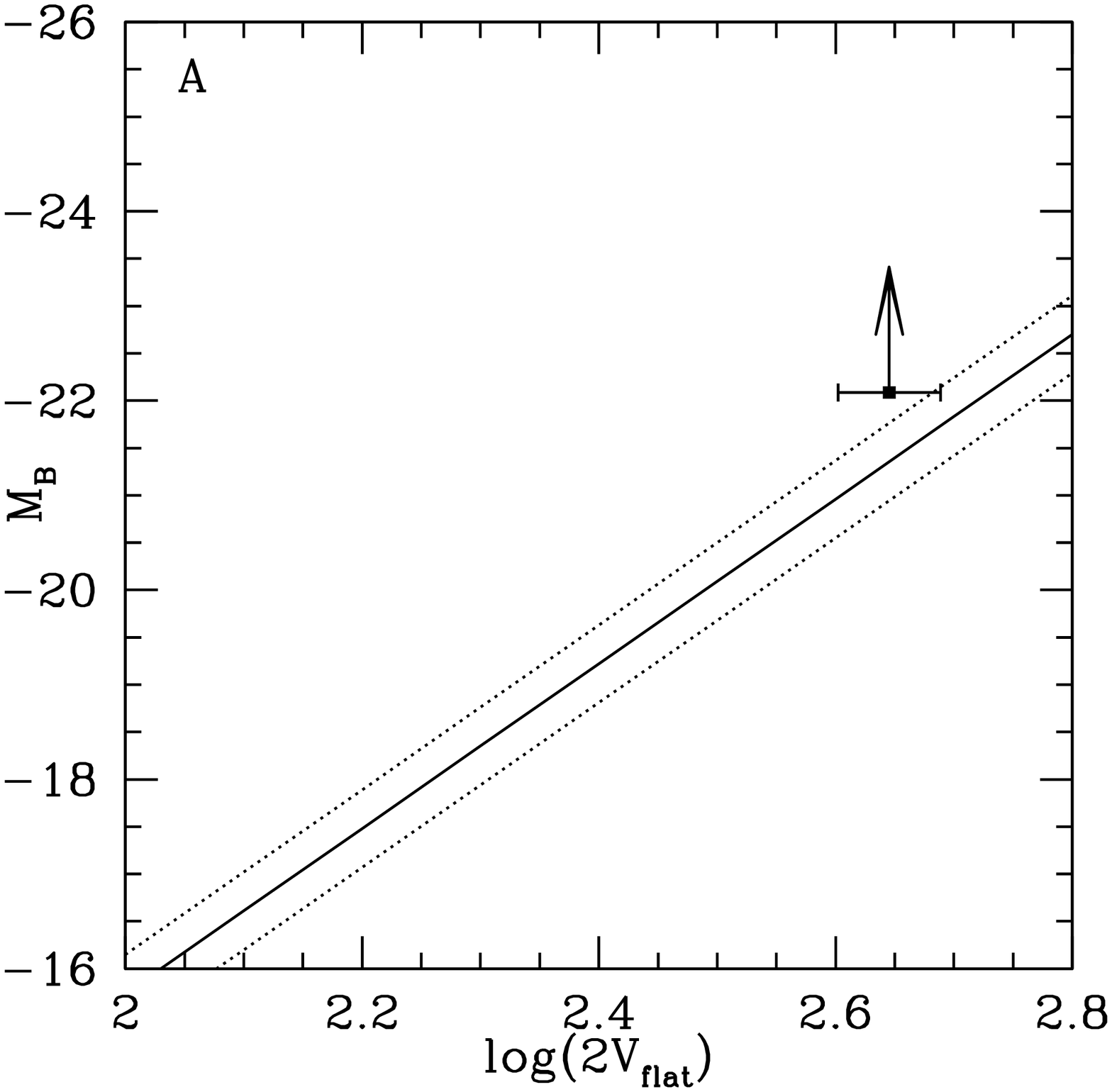}
\includegraphics[width=9cm]{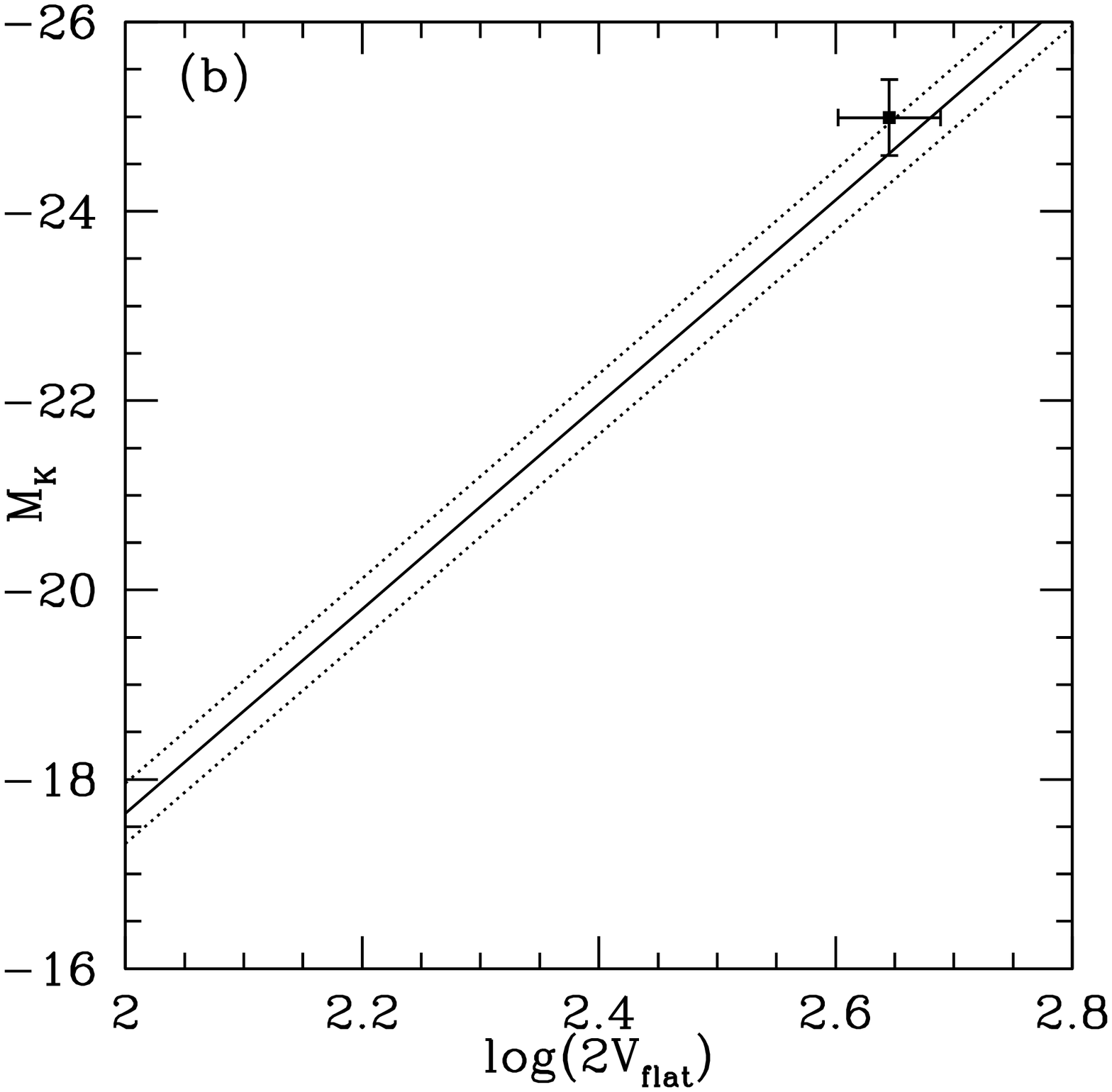}

   \caption{The rest frame B band (left) and rest frame K band (right) TFR. The arrow indicates the extinction correction using the best fit $A_V$ from Wuyts et al. (\cite{wuyts}). The local TFR and its scatter (full and dashed lines) are from Verheijen (\cite{verheijen}). The errorbar on the absolute B magnitude is smaller than the point size and is therefore not shown. }
       \label{tfr}
\end{figure*}

In Fig. \ref{tfr} we plot the local rest frame B band and rest frame K band TFR of Verheijen (\cite{verheijen}) together with our data point for F257. F257 is 
marginally brighter than expected from the local B band TFR. The offset is $0.7 \pm 0.4$ mag in B band, where the errorbar reflects the uncertainty due to the velocity uncertainty. Correcting for extinction using the $A_V$ found by Wuyts et al. (\cite{wuyts}), the offset increases to 2.1 mag. For comparison: the $1\sigma$ dispersion in the B band TFR as found by Verheijen (\cite{verheijen}) is 0.41 measured along the magnitude axis.

F257 is consistent with the local K band TFR with  $0.4 \pm 0.4 \pm 0.5$ mag brightening where the first errorbar reflects the errors bars in the IRAC fluxes from which the rest frame K band magnitude was calculated (F257 is blended with a nearby source), and the second reflects the uncertainty due to the velocity uncertainty. For comparison: the $1\sigma$ dispersion in the K band TFR as found by Verheijen (\cite{verheijen}) is 0.32.

Local TFRs differ slightly from each other due to for example different choice of samples, different velocity parameters (e.g. linewidth or velocity, and also maximum or flat RC velocity) and RC fitting methods (e.g. Puech et al. \cite{puech8}, Kannappan \& Barton \cite{kannappanbarton}). The differences are smallest at the high mass end of the TFR and do not change our conclusions (van Starkenburg et al. \cite{vanstarkenburg} and references therein). We choose the Verheijen (\cite{verheijen}) TFR as reference TFR because it is based on velocity fields, it is defined in optical and near-IR bands and uses the flat RC velocity, which (locally) minimizes the scatter in the TFR.

\subsection{The stellar and baryonic mass TFR}
\begin{figure*}
   \centering
\includegraphics[width=9cm]{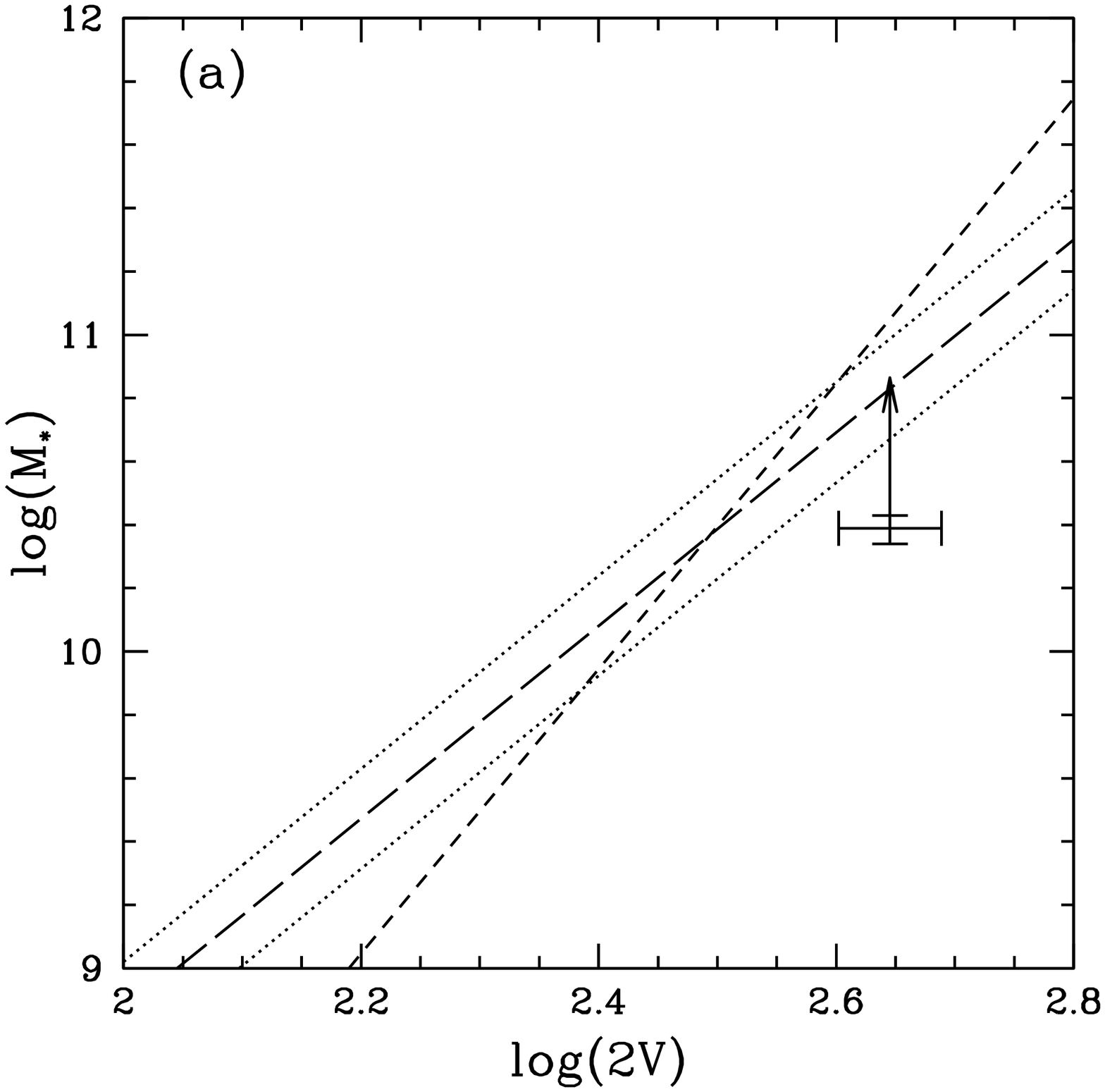}
\includegraphics[width=9cm]{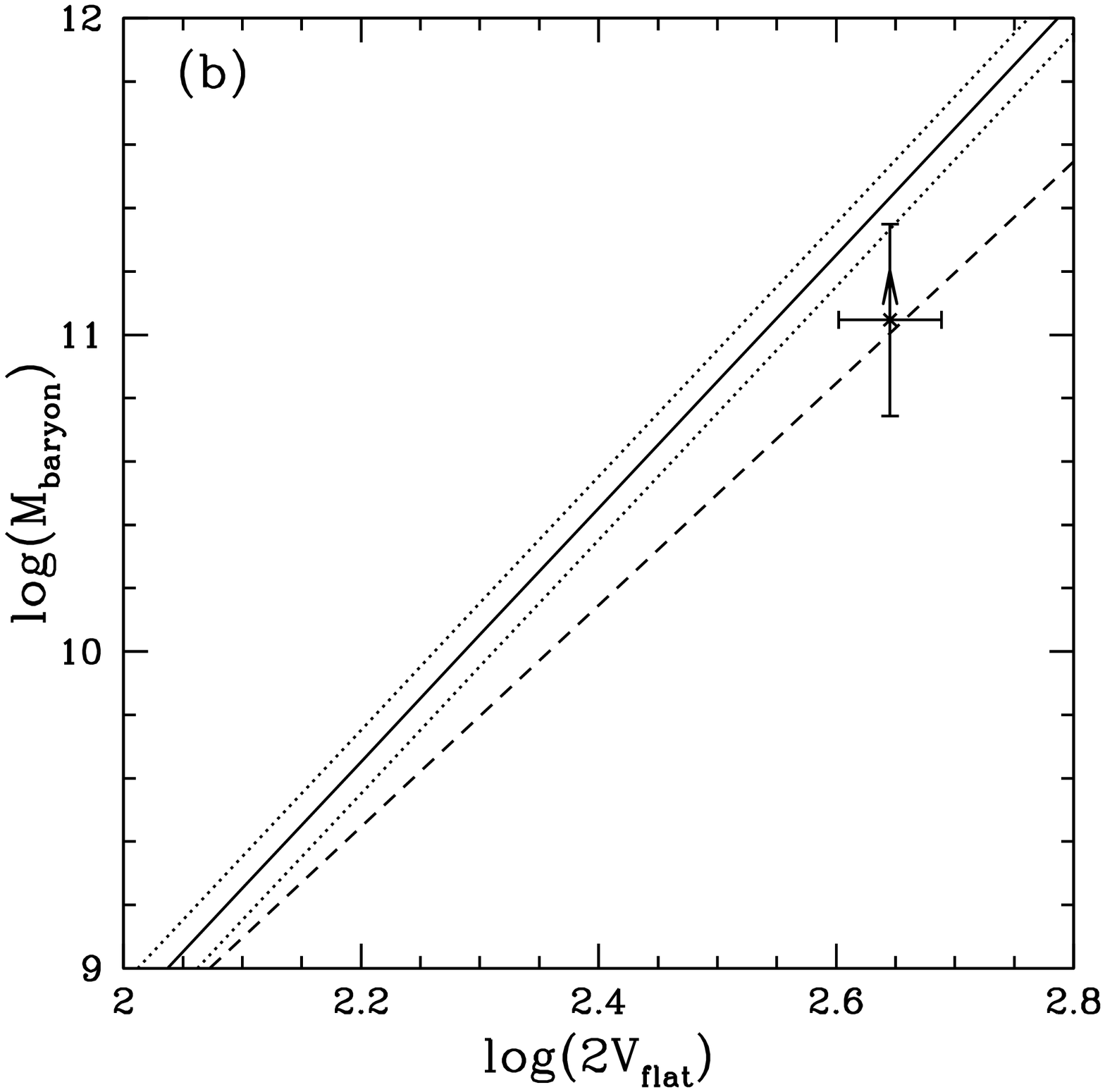}

   \caption{a. The local stellar mass TFR and $1\sigma$ scatter of Pizagno et al. (\cite{pizagno}) (long dashed and dotted lines) and Bell \& de Jong (\cite{bell}) (dashed line). b. The local baryonic mass TFR and $1\sigma$ scatter of Verheijen (\cite{verheijen}) and Bell \& de Jong (\cite{bell}) (dashed line).  The arrows indicate where F257 would lie for the maximally old plus young burst assumption (see text). }
       \label{sbtfr}
\end{figure*}

The relation between the stellar mass and rotation velocity is known as the stellar mass TFR. When we compare F257 with the local stellar mass TFRs of Bell \& de Jong (\cite{bell}) and Pizagno et al. (\cite{pizagno}) as shown in Fig. \ref{sbtfr}a, we find that F257 lies below these stellar mass TFRs (after reducing the stellar mass found by Wuyts et al. \cite{wuyts} by 30\% to account for the different IMFs used). The offset in $\log(M_*)$ is $0.66 \pm 0.20 $ and $0.44 \pm 0.12$ respectively where the errorbars reflect the uncertainty due to the uncertainty in the velocity (the uncertainty in the stellar mass was 0.05). The $1\sigma$ dispersion in the Pizagno et al. stellar mass TFR is 0.158 so the offset is significant at the $\sim 2 \sigma$ level.

Wuyts et al. (\cite{wuyts}) also discuss the effect on the best fit stellar mass of fitting a 2-component model instead of a 1-component model to the SED. The best fit stellar mass for a maximally old population plus a 100 Myr burst for F257 is $1.0 \times 10^{11} \textrm{ M}_\odot$. This brings F257 in agreement with the local stellar mass TFR, as indicated by the arrow in Fig. \ref{sbtfr}. It is interesting to note that this 2-component model mass makes F257 a nearly ($\sim 75\%$) maximum disk.

Some authors (e.g. Verheijen \cite{verheijen}, Bell \& de Jong \cite{bell}) added the gas mass to the stellar mass to get a baryonic mass TFR, see also Fig. \ref{sbtfr}b. Although the gas mass of F257 is uncertain due to the large scatter in the Kennicutt relation, we also compare F257 with these local baryonic mass TFRs and find that F257 is consistent with the baryonic mass TFR from Bell \& de Jong (\cite{bell}): $\log(M_{bar}) =-0.04 \pm 0.29 \pm 0.3$ where the first errorbar reflects the uncertainty due to the uncertainty in the velocity and the second errorbar reflects the uncertainty in the baryonic mass (assuming of a factor 2 uncertainty in the gas mass). 

Although Bell \& de Jong (\cite{bell}) use an earlier version of the data set of Verheijen (\cite{verheijen}), they find significantly different baryonic mass TFRs, especially at the high mass end. This difference is explained by the different $M/L$ that they use to convert K band light to stellar mass. The (constant) $M/L$ used by Verheijen is $\sim 3 $ times larger than the (best fit for each galaxy) $M/L$ of Bell \& de Jong (\cite{bell}). Verheijen (\cite{verheijen}) uses this $M/L$ because it minimizes the scatter in the baryonic mass TFR. However, according to Bell \& de Jong, this $M/L$ is unrealistically high. A lower $M/L$ would shift the Verheijen relation downward, especially at the high mass (i.e. low gas fraction) end.

\subsection{Comparison to other high-$z$ samples}

Several authors have attempted to measure rest frame near-IR and stellar mass TFRs at high redshift. 
Conselice et al. (\cite{conselice}) analyzed 101 disk galaxies in the redshift range $0.2 < z < 1.2$ and find no evolution in the rest frame K band TFR and in the stellar mass TFR.

Kassin et al. (\cite{kassin}) use a combination of rotation velocities and velocity dispersions ($S_{0.5} = \sqrt{0.5V^{2}_{rot}+\sigma^{2}_{gas}}$) to construct a stellar mass TFR for galaxies with strong emission lines in the redshift range $z=0.1-1.2$. This reduces the scatter in the stellar mass TFR. However, they find no $z$-evolution in the stellar mass TFR using this $S_{0.5}$ parameter. F257 is a rotation dominated galaxy and hence consistent with their normal stellar mass TFR and their $S_{0.5}$ parameter stellar mass TFR.

Weiner et al. (\cite{weiner}) study the rest frame B and J band TFR out to $z=1.2$. They find significant brightening in rest frame B band (1-1.5 mag between redshift 0.4 and 1.2). However, although they find that the slope of the J band TFR evolves, the overall luminosity in J band does not.

Flores et al. (\cite{flores}) published the first high redshift TFR using integral field spectroscopy. They observed a sample of [\ion{O}{ii}] emitting galaxies at $z=0.4-0.75$ using the GIRAFFE IFU on the VLT. They find that only 35\% of their sample are rotating disks in their definition (the major axis of the velocity field coincides with the major axis of photometry plus a peak in centre of the dispersion field). They find no evolution and small scatter in the rest frame K band and stellar mass TFR and some galaxies that are brighter in rest frame B band.

To summarize, other authors have not found evolution in the near-IR or the stellar mass TFR out to redshift 1.2. This non-evolution in near-IR $z=1$ TFR was predicted by Buchalter, Jimenez \& Kamionkowski (\cite{buchalter}) and in the stellar mass $z=1$ TFR by Portinari \& Sommer-Larsen (\cite{portinari}). Our findings suggest that the K band TFR may not evolve out to redshift 2. F257 is $\sim 2 \sigma$ below the stellar mass TFR, a hint that the stellar mass of F257 is still building up, which would be consistent with its young age, large SFR and gas mass.
However, it should be noted that the offset depends on the SED modeling. If a maximally old population is added, F257 is consistent with the local stellar mass TFR.

\subsection{Summary and discussion}

We conclude that F257 is a fully formed, rotationally supported disk galaxy. Within the errorbars, it is consistent with the local K band TFR. The small offset from the local B band TFR can be explained by star formation activity and the young stellar population. 

F257 lies slightly below the local stellar mass TFR, an offset significant at the $2\sigma$ level. Evolving from redshift 2 to the present, the stellar mass must increase without a significant increase in the K band luminosity. However, the best fit stellar mass strongly depends on the SED modeling. When we add a maximally old population, F257 is in agreement with the local stellar mass TFR.

 F257 is consistent with the baryonic mass TFR, but the uncertainties are large. 
To study offsets from the baryonic TFR, one needs more galaxies and/or more accurate gas masses. We present our full sample in van Starkenburg et al. (\cite{prep2}). In the near future, ALMA will provide us with the latter.

We finally note that the regular dynamcial sctructure of F257 is remarkable for its young age: the estimated age of the stellar population of $\sim 160 Myr$ equals the rotation period of the galaxy at $1 - 2 R_d$ and at $1-2 R_{flat,true}$). Therefore, the galaxy has dynamically relaxed on a very short timescale, much shorter than the timescale for the buildup of the final stellar population.

\section{Summary and conclusions}

Labb\'e et al. (\cite{labbe}) presented photometric evidence that F257 is a large disk galaxy at $z=2.03$. We confirm the disk nature of this galaxy by its kinematical properties and find that is in many respects similar to local disk galaxies. Specifically, we find:
\begin{enumerate}
 \item The velocity field is consistent with a rotating disk galaxy, a tilted ring model leaves small random residuals and no evidence for non circular motions;
 \item There is no evidence for an AGN in the form of broad lines or a large [\ion{N}{ii}]/H$\alpha$ ratio;
 \item The [\ion{N}{ii}]/H$\alpha$ ratio peaks in the centre; 
 \item The disk is primarily rotation supported: $V/\sigma_z\sim3.9$; 
 \item The dispersion increases towards the center;
 \item The RC slowly rises like that of a late-type spiral in the local universe and then flattens. There is no evidence for dark matter within $r < 12 \textrm{kpc}$;
 \item F257 has a specific angular momentum and spin parameter comparable to those of local galaxies;
 \item F257 is consistent with the local K band TFR;
 \item F257 lies below the local stellar mass TFR at the $2\sigma$ level and is consistent with the local baryonic mass TFR. A 2-component model for the SFH increases the stellar mass so that F257 is consistent with the stellar and baryonic mass TFR and makes F257 a nearly maximum disk.

\end{enumerate}
Despite its very young age, high star formation rate, gas mass and clumpy star formation, F257 is very similar to local late type spiral galaxies. Its dynamical properties are more like those of local galaxies than those of other galaxies observed at similar redshift. 

These findings put constraints on the evolution of F257 from $z=2.03$ to the present. If the stellar mass increases so that F257 becomes consistent with the local stellar mass TFR, the K band mass-to-light ratio must also increase so that that the total K luminosity does not change. Infalling gas must have the same specific angular momentum as the galaxy, i.e. it must fall in from the outer parts of the present halo.

\begin{acknowledgements}
LvS wishes to thank Juha Reunanen for lengthy discussions about SINFONI data reduction. We thank Eric Bell for an interesting discussion and Marc Verheijen for his help with the local TFRs.

\end{acknowledgements}

\begin{appendix}

\section{FOV reconstruction}\label{app}

\subsection{Introduction}

Our targets are too faint to acquire directly or to detect in individual frames. When we made our first observations with SINFONI, we adopted the normal strategy for these faint targets: blind offsetting from a nearby bright star. We also followed the recommendation in the SINFONI manual to reobserve the blind offset star in the same OB as the science observations to be able to accurately combine the data from different OBs. We refer to these star observations as "PSF stars" (the template used for the observation is the calibration template ``PSF'').

\begin{figure}
  \resizebox{\hsize}{!}{\includegraphics{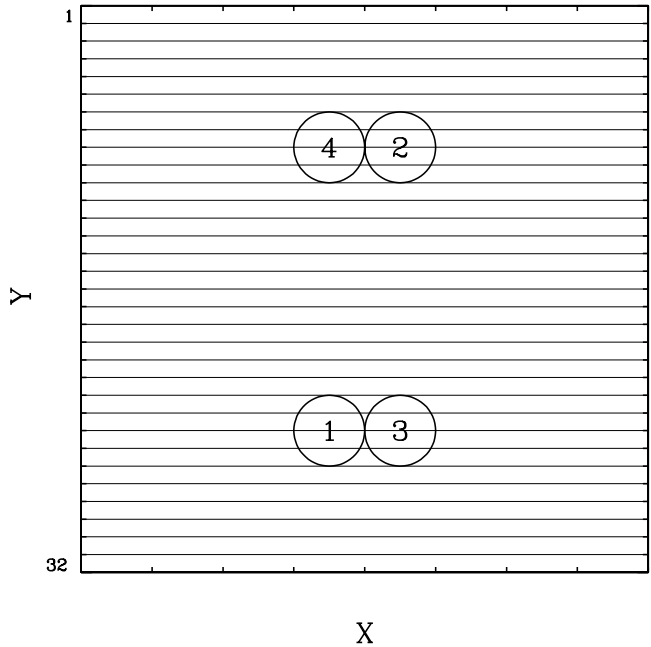}}
 \caption{SINFONI observing strategy. The target is observed at positions 1, 2, 3 and 4. The distance between position 1 and 3 is 1\arcsec, between 1 and 4 4\arcsec. The positions of the slitlets and the X and Y directions are marked. Slitlets are numbered from top to bottom. The total FOV size is 8\arcsec times 8\arcsec.}
  \label{obs_mariska}
\end{figure}

We also followed the standard procedure to reconstruct the FOV from the raw images. SINFONI's FOV is sliced into 32 slitlets (see also Fig. \ref{obs_mariska}) that are combined into a single pseudo-slit and then dispersed. These 32 slitlets are projected on the detector in a brick-wall pattern. The slitlet distances (measured on calibration data made with fibers) define the distances on the detector between points with the same x coordinate (see Fig. \ref{obs_mariska} for a definition of the x and y direction). These slitlet distances are used to reconstruct the FOV in the reduction process. 

\begin{figure}
  \resizebox{\hsize}{!}{\includegraphics{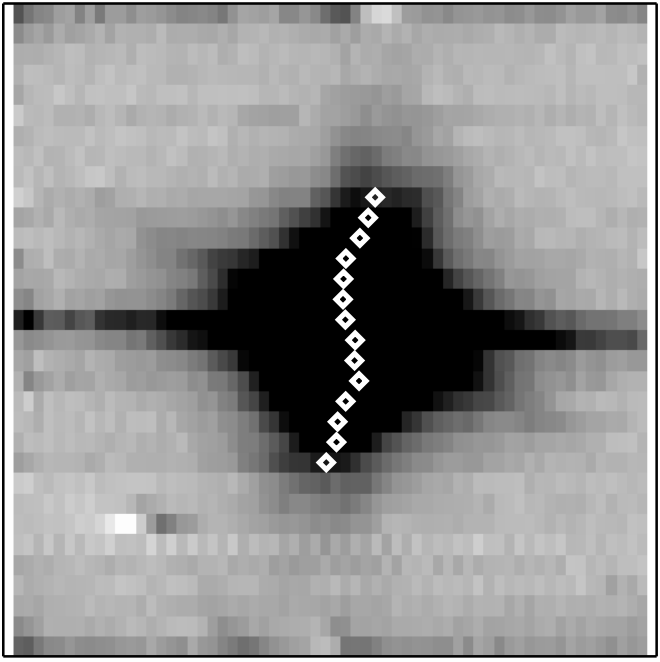}}
 \caption{The reconstructed image of one of the bright standard stars observed for F257. To guide the eye, the center of the star in each slitlet has been marked. }
  \label{shape}
\end{figure}

Having reduced our first data sets, we noticed that the reconstructed images of the PSF and standard stars showed some peculiarities. The centre of the star changed from slitlet to slitlet, often back and forth, but in some cases the gradual shift in centre resulted in a stretched stellar image. An example is shown in Fig. \ref{shape}. Also, the overall shape of the stars was not round but elliptical, especially if the seeing was good. FWHM ratios varied between 1 and 1.6, much larger than we could explain. As all reconstructed stars looked different, a simple error in the slitlet distances was excluded. 

After careful study of the reconstructed images of our PSF and standard stars of our first observing runs, we concluded that the details of SINFONI's FOV reconstruction are more complicated and that the quality of the velocity fields measured might be improved if the details are treated carefully. In this Appendix, we describe our investigations, the effects we found, their effect on the data quality and our new improved observing strategy.

\subsection{Further investigations: varying pixel scale}

To investigate the details of SINFONI's FOV reconstruction, a PSF star was observed 
at the same positions in the FOV as the science target, in some cases even before and after the science target. The positions in the SINFONI FOV are shown in Fig. \ref{obs_mariska}.

\begin{figure}
  \resizebox{\hsize}{!}{\includegraphics{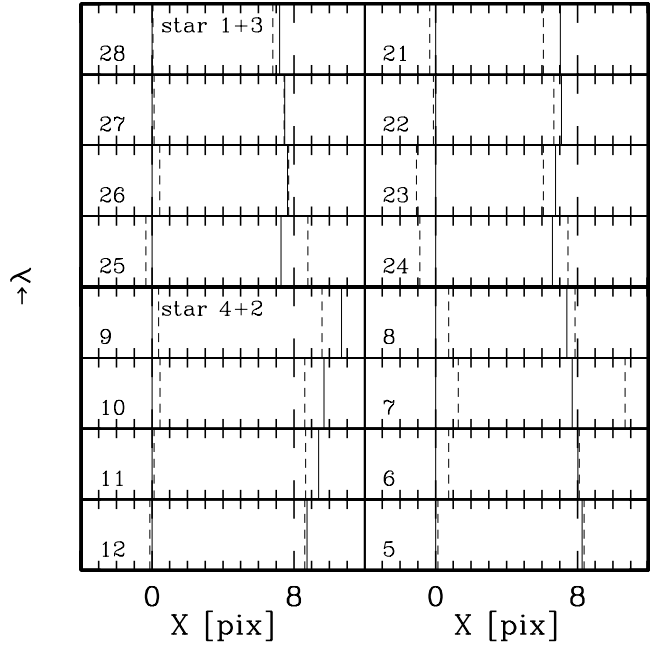}}
 \caption{The star traces for the test observing sequence with observing strategy as shown in Fig. \ref{obs_mariska}.  Each panel represents one slitlet, the slitlet number is indicated in the lower left of each panel. The vertical lines show the positions of the star traces before (full lines) and after (dashed) lines observations of the science frames (time difference is about 1 hour). For each slitlet, the position of the first observation of the left star has been shifted to zero.  The centers of the two stars could be measured in 8 slitlets. Those of the two stars in the lower half of the FOV (stars 1 and 3 in Fig. \ref{obs_mariska}) are in the upper 8 panels, those of the stars in the upper half of the FOV (stars 2 and 4 in Fig. \ref{obs_mariska}) are in the lower 8 panels.
The distance between two star traces should be 8 pixels if the pixel scale is 0\farcs125. Note that the distances between the two stars are different and change differently in each slitlet. Also note that the distance is less than 8 pixels for most slitlets in the lower half of the FOV and more than 8 pixels for most slitlets in the upper half of the FOV.   }
  \label{scalevar}
\end{figure}

When we compared the star traces of the PSF stars that are in the same slitlets in the {\it raw} frames (i.e. stars 1 and 3, and stars 2 and 4 in Fig. \ref{obs_mariska}), we noted that the distance between the star traces varies as is illustrated in Fig. \ref{scalevar}. From the telescope offset in the x direction (1\arcsec) and SINFONI's spatial pixel scale ($0\farcs125 \textrm{ pix}^{-1}$) the distance between the star traces should be 8 pixels in all slitlets. The average value we found was indeed 8 pixels, but we found offsets of $\sim 10\%$ around this value (see Fig. \ref{scalevar}). In other words: the pixel scale varied up to 10\%. The effect was not stable in time, we have seen significant changes in only 1 hour as can also been seen in Fig. \ref{scalevar}. 

We emphasize that we measured these positions on the {\it raw} frames, so that reduction artifacts are excluded. Also, this effect is not caused by errors in the telescope offset: the pixel scale also varied (by less than 10\%) from slitlet to slitlet for a single pair (see the upper or lower half of the panels in Fig. \ref{scalevar}), and the pixel scale in the y direction (i.e. the distance between stars 1 and 4 or 2 and 3 in Fig. \ref{obs_mariska}) was stable to measurable accuracy.

This explains the odd shapes of the reconstructed stellar images. The slitlet distances measured on fiber data are only valid for that observation and only approximately valid for other observations. The changing pixel scale explains why the FWHM of the stars in x and y direction is not the same for good seeing data.

SINFONI is a Cassegrain-mounted instrument, which makes is sensitive to flexure. As we saw in later observing runs (where we adopted our new multiple PSF star strategy (see below)) in many consecutive hours of observations of the same target, the pixel scale changes smoothly in time and hence parallactic angle. Therefore, we suspect flexure in the instrument as the origin of the effects described above. This could also explain the 'moving slitlets' phenomenon: for some parallactic angles, the positions of the slitlets on the detector shifts up to $\sim 2$ pixels in a few hours (of observing the same target).

We do not have sufficient observations of PSF stars to be able to predict the pixel scale for a given slitlet at a given target at a given time (or even to be sure that pixel scale can be predicted from parameters like parallactic angle and airmass). We have indications that the upper and lower half of the FOV have systematically different pixel scales (as is the case for the example shown in Fig. \ref{scalevar}). We only looked at data taken with SINFONI's largest FOV  without AO and have not searched for differences between J, H, H+K and K band although we have noted these effects in all 4 filters.

\subsection{Effects on data quality}

Given these complex pixel scale changes, the question arose what the consequences were for data sets with observation of a single PSF star and what improvements could be made in future observations (including the observations discussed in this paper). 

We assumed that the addition of many frames will average out the pixel scale differences so that the pixel scale in the combined image is the same in both spatial directions: 0\farcs125 (which seems safe to assume, as the pixel scales we find have on average this value). The changes in object size are negligible compared to the seeing even for the large high redshift galaxies in our sample, moreover seeing variations are usually larger than 10\%. 

However, the varying pixel scale results in a position uncertainty which can be important. In our science exposures, the telescope offset between A and B frames in the x direction is about 4\arcsec~ for all our data (this paper, and van Starkenburg et al. \cite {prep1}, \cite{prep2}). 10\% of 4\arcsec~ is comparable to our best seeing observations (e.g. this paper 0\farcs55). This position uncertainty causes an additional smear in the x direction which cannot be removed with the observation of one PSF star per OB. As our targets are too faint to measure their position in individual frames, our new observing strategy should solve this position uncertainty in another way. 

To summarize: the varying pixel scale causes a position uncertainty that could degrade the data quality in one spatial direction if the target is faint and observed under good seeing conditions.

\subsection{Multiple PSF star observing strategy}

Given these results, we changed our observing strategy as follows. The position uncertainty is the most important factor to correct. Observing a PSF star at every position where we observed the science target and as closely in time as possible provides us with accurate positions (i.e. in term of an ABB'A'-like observing cycle $\textrm{A}_\textrm{psf}\textrm{B}_\textrm{psf}\textrm{A}_\textrm{science}\textrm{B}_\textrm{science} \textrm{A'}_\textrm{science}\textrm{B'}_\textrm{science}    \textrm{A'}_\textrm{psf}\textrm{B'}_\textrm{psf}$). We then use these PSF positions when we combine our data. 

As a bonus, we can seeing-weight the images (using the FWHM in the y direction) to improve image quality further. Another benefit is that we do not have to worry about the 'moving slitlet' effect anymore (it is unclear whether the moving slitlets are simply a shift in projection of the same part of the sky or a projection of a slightly different part of the sky or a combination of both). 

When preparing the observations, we also take into account that the FOV size might be slightly smaller in the x direction than expected or that the target might be closer to the edge of the FOV than you would expect from the offsets. Also, when possible we put the major axis of the galaxy along the y direction in the FOV to ensure maxium resolution where we need it most. 

We did not attempt to correct the individual frames and slitlets for the pixel scale variations as we saw before that these effects are negligible compared to the seeing. Moreover, in the case of correcting individual slitlets, one would need very bright PSF stars to get enough flux in all slitlets where the (large) science target is, and those are in general not available (the example shown in Fig. \ref{scalevar} is with a very bright PSF star). 

\subsection{Other artifacts}

For the very bright standard stars, we noted that their centres in the reconstructed image varied as a function of wavelength. This may indicate even more complex flexure effects, but a more probable explanation is that the distortion correction of the (earlier versions of the) SINFONI pipeline did not remove the distortion entirely. As we are only interested in a single emission line, we solved this by measuring the position and shape of the PSF stars at the wavelength of the emission line. 

\end{appendix}

\end{document}